\documentclass[12pt, preprint]{aastex}

\usepackage{graphicx}
\usepackage{natbib}
\usepackage{epstopdf}

\def\gapp{\lower 3pt\hbox{${\buildrel > \over \sim}$}\ }
\def\lapp{\lower 3pt\hbox{${\buildrel < \over \sim}$}\ }

\slugcomment{To appear in The Astrophysical Journal}

\shorttitle{Multi-Epoch Observations of  HD69830}

\shortauthors{Beichman et al.}

\graphicspath{{converted_graphics/}}
\begin{document}

\title{Multi-Epoch Observations of HD69830: \\
High Resolution Spectroscopy and Limits to Variability}

\author{
C. A. Beichman$^{1}$, 
C. M. Lisse$^2$,
A. M. Tanner$^{1,3}$, 
G. Bryden$^{1,4}$, 
R. L. Akeson$^{1}$, \\
D. R. Ciardi$^{1}$, 
A. F. Boden$^5$,
S. E. Dodson-Robinson$^{6}$,
C. Salyk$^{6}$, and
M.~C. Wyatt$^{7}$
}
\affil{1) NASA Exoplanet Science Institute, Jet Propulsion Laboratory, California Institute of Technology, M/S 100-22, Pasadena, CA, 91125}
\affil{2) Johns Hopkins University, Applied Physics Laboratory, 
  11100 Johns Hopkins Rd, Laurel, MD 20723}
\affil{3) Department of Physics and
Astronomy, MSU, MS, 39762}
\affil{4) Jet Propulsion Laboratory, California Institute of Technology, 
  4800 Oak Grove Dr, Pasadena, CA 91109} 
\affil{5) Caltech Optical Observatories, California Institute of Technology, M/S 100-22, Pasadena, CA, 91125}
\affil{6) University of Texas, Astronomy Department, 1 University Station 
C1400, Austin, TX 78712}
\affil{7) Institute of Astronomy, University of Cambridge, 
  Cambridge, CB3 0HA, UK}

\email{chas@pop.jpl.nasa.gov}

\begin{abstract}

The main-sequence solar-type star HD69830 has an unusually large amount of dusty debris orbiting close to three  planets found via the radial velocity technique. In order to explore the dynamical interaction between the dust
and planets, we have performed multi-epoch photometry and spectroscopy of the system over several orbits of the outer dust. We find no evidence for changes in either the dust amount or its composition, with  upper limits of 5-7\% (1 $\sigma$ per spectral element) on the variability of the {\it dust spectrum} over 1 year, 3.3\% (1 $\sigma$) on the broad-band disk emission over 4 years, and 33\% (1 $\sigma$) on the broad-band disk emission over 24 years. Detailed modeling of the spectrum of the  emitting dust  indicates that the dust is located outside of the orbits of the three planets and has a composition similar to  main-belt, C-type asteroids asteroids in our solar system. Additionally, we find no evidence for a wide variety of gas species associated with the dust. Our new higher SNR spectra do not confirm our previously claimed detection of H$_2$O ice leading to a firm conclusion that the debris can be associated with the break-up of one or more C-type asteroids  formed  in the dry, inner regions of the protoplanetary disk of the HD69830 system. The modeling of the spectral energy distribution and high spatial resolution observations in the mid-infrared are consistent with a $\sim$ 1 AU location for the emitting material.

\end{abstract}


\section{Introduction}

One of  Spitzer's most intriguing discoveries is the extreme level of zodiacal emission around the nearby (12.6 pc)  K0V star HD69830. The dust cloud around HD69830 is approximately 1,400 times brighter than the  emission of our own zodiacal cloud and shows a plethora of solid state features attributable to small, hot, crystalline silicate grains located $<$1 AU from the parent star \citep{beichman05irs}. Such intense emission from dust in the inner solar system is exceedingly rare, detectable in only $\sim$1\% of the  mature stars surveyed by Spitzer \citep{beichman06tpf, trilling08, lawler09}. Interest in this cloud and its link to the evolution of planetary systems was greatly heightened by the discovery of three Neptune-mass planets orbiting within 0.6 AU of the star \citep{lovis06}. One of the major unanswered questions about this system is whether the dust seen by Spitzer comes from collisions within a particularly massive  asteroid belt  or from a swarm of comets released by planet-disk interactions. The Spitzer follow-up observations described here include high signal to noise (SNR), low and high spectral resolution observations of the dust disk to help distinguish between these two alternatives. Five repeats of the spectral observations over 12  months were designed to search for small variations that might be expected from a dust cloud evolving on a dynamical time scale of less than 1 year. Overall, the data described here span more than 4 years of Spitzer observations from 2004 to 2008 and extend as far back as a 1983 detection by IRAS at 25 $\mu$m. In this paper we use new Spitzer data to refine our assay of  the mineralogical and gaseous components of the HD69830 disk \citep{lisse07} and look for temporal variations in the disk emission that might occur on dynamical  or dust replenishment timescales, from 1 year to over 1,000 years. Observations of young protoplanetary disks can exhibit dramatic changes in the shape of their Spitzer/IRS spectra on timescales as short as one week \citep{muzerolle09}. Our observations were designed to probe whether such activity is present in the diffuse debris disk orbiting a much older main sequence star.

To accomplish these aims, we made observations using all three Spitzer instruments (\S\ref{spitObs}). The location of the emitting material in relation to the three radial velocity (RV) planets is also of considerable interest in understanding the origin of the dust which might lie either interior to, in-between or exterior to the planets. To investigate this question we made mid-IR images with the Michelle mid-IR camera on the 8-m Gemini telescope and looked for extended near-IR emission with the 85-m baseline Keck Interferometer (\S\ref{ground}). After a discussion of the entire set of observations, we describe our results in \S\ref{results} and discuss their implications in \S\ref{discuss}.

\section{Observations }

\subsection{Spitzer Observations and Data Reduction}\label{spitObs}

Observations were made of HD69830 using all three Spitzer instruments (Table~\ref{ObsLog}) --- the IRAC \& MIPS cameras  and the IRS spectrograph  \citep{fazio2004, rieke2004, Houck04}. In addition to discovery data obtained with MIPS and IRS in 2004, the two photometric instruments were used at one epoch in 2007 while the IRS spectrometer was used on 5 occasions at low resolution and 6 occasions at high resolution to look for small variations in the emission from small dust grains responsible for the excess from this source. In the case of the IRS observations, a nearby star HD68146 (F7V) which is known to have no long wavelength excess \citep{beichman06tpf} was used as a reference star for both flat fielding and bad pixel monitoring. In addition, the IRS peak-up array was used to obtain 22 $\mu$m photometry along with each IRS spectrum. We have used the IRS peak-up photometry to look for temporal variations and in the absence of any variability, we averaged the IRS spectral data to obtain the best spectrum of the excess for comparison with dust models. We also examined the spectral data for evidence for temporal variations.

\subsubsection{IRAC}

IRAC images were obtained at all four wavelengths from 3.6 to 7.9 $\mu$m. To avoid saturation by the bright stars considered here, 
short 0.02-sec exposures were made in the sub-array mode. Co-adding 64 individual sub-array frames, we performed aperture 
photometry using standard aperture size, aperture correction, and calibration factors as described in the {\it Operating Manual}\footnote{http://irsa.ipac.caltech.edu/data/SPITZER/docs/irac/iracinstrumenthandbook/}, yielding the values listed in Table~\ref{phottable}. 
The photometric uncertainty is dominated by the overall IRAC calibration accuracy of 3\% \citep{reach05}. The observed fluxes in conjunction with ground-based data constrain the photospheric emission from the star.  No evidence for an excess is seen in any of the four bands which puts a strong constraint on the amount of excess  at the short end of the IRS spectra,  wavelengths $\lapp 8\mu$m in modules SL1 and SL2 (\S\ref{lores} below).

\subsubsection{MIPS}\label{mips}

MIPS observations were made at 24 and 70 $\mu$m with an integration time at 70 $\mu$m  significantly greater than in the 
initial survey \citep{bryden06}, 2200 sec vs. 340 sec. Data were reduced using the MIPS instrument team pipeline version 3.10
with standard apertures and calibration values  \citep{engelbracht07, gordon07}. The new 24  $\mu$m and 70 $\mu$m fluxes are consistent with previous measurements. At 24 $\mu$m, the results are identical within the calibration uncertainty of 2\%. At  70 $\mu$m we measured $19\pm4$ mJy \citep{bryden06} and in the new data reported here, $15\pm 3$ mJy. We adopt the new MIPS 70 \micron\ observation with its greater number of scans and longer integration time as the more reliable.  As expected from the original observations, there is a very strong infrared excess at 24 $\mu$m ($\sim$50\% over the predicted photosphere). However, despite the longer integration at 70 $\mu$m there remains  no  evidence for a long wavelength excess. Any 70 $\mu$m excess  must be below our confusion-limited detection limit, $<$50\% ($3\sigma$), over the photosphere.

\subsubsection{IRS Low Resolution}\label{lores}

Low-resolution IRS measurements were made in all four modules  (SL2, SL1, LL2, LL1) covering wavelengths from 5 to 35 $\mu$m
\citep{Houck04}. A high precision peak-up was performed to ensure careful centering of the star within the slits. 
Significantly greater integration times were used for the LL1 and LL2 measurements  than were allotted in the initial IRS survey \citep{beichman05irs},
increasing the sensitivity to weak features at long wavelengths.  New observations reported in this paper include 3 repeats spaced over 1 year (Dec 2007, Apr 2008, and Dec 2008) plus two additional repeats granted as part of the Director's Discretionary Time (DDT, both in Jan 2009).

The data were reduced using standard techniques of subtracting the spectra obtained in the two Nod positions  from  one another for sky subtraction. Uncertainties at each wavelength are based on the differences between  the two Nod positions.  To derive flat field corrections we obtained contemporaneous IRS observations on three occasions (Repeats 1-3 in Table~\ref{ObsLog}) of the reference star HD68146, an F7V star of comparable magnitude and no known IR excess located a few degrees away from HD69830.  At the longest wavelengths, corrections to the HD69830 data  based on observations of HD68146 relative to its predicted photosphere (based on a Kurucz model as described in Beichman et al. 2006a) proved to be  no better than simply adopting the average flat field calibration  provided by the Spitzer Science Center. 
Thus, no additional corrections were  made to the HD69830 data in the LL1 or LL2 modules; only the data from the shortest wavelength modules (SL1 and SL2) 
were corrected for small flat field errors using contemporaneous HD68146 measurements.  This correction is particularly important for correcting
the known ``tear-drop" artifact at wavelengths  between 12-14 $\mu$m in the SL1 module\footnote{http://irsa.ipac.caltech.edu/data/SPITZER/docs/irs/irsinstrumenthandbook/78}. While the large scale concavity  introduced into the spectrum is modest, about 5\% in total flux, its effect on this work is  amplified by subtraction of the stellar photosphere. 
The shorter wavelengths in  the spectrum are  unaffected. The amplitude and shape of the ``tear-drop'' changed from scan to scan and  on timescales as short as the time required to move from the reference star to HD69830, consistent with the hypothesis that the tear-drop originates from small centering errors of the star in the spectrometer slit. Overall, however, the large scale shape of the tear-drop remained relatively consistent between the 3 epochs that have corresponding control star observations of HD68146, such that a flat-field based on the 3-epoch average  removed much of the tear-drop curvature from both stars' spectra. As discussed below, however, the residuals from the tear-drop make it difficult to confirm  the broad, weak spectral feature  due to water vapor first noted in \citet{lisse07}.

Additional calibration corrections were made to ensure continuity between the orders of each spectrum. At the shortest wavelengths of SL1 the spectra are  consistent with the photospheric model of HD69830 (Beichman et al. 2005 and $\S$3.1) which was in turn linked to ground-based and IRAC photometry; no scaling of SL1 is required. SL2 data are similarly consistent with the photosphere and do not require any scaling. 
For the longest wavelengths, LL1 is scaled to match the MIPS 24 $\mu$m flux. The correction factors are very small: +0.5\%, +2.2\%, and $-$0.2\%
for Repeat1, Repeat2 and Repeat3, respectively. Intermediate wavelengths (order LL2) are most problematic  in that they exhibit the most variation between epochs and require the largest corrections in order to properly line up with the bordering SL1 and LL1 data.
The LL2 correction factors for the 3 epochs are $-$5\% $-$3\%, and $-$6\% for Repeat1, Repeat2 and Repeat3, respectively.

After first verifying that there were no statistically significant variations that could not be attributed to either noise or instrumental effects ($\S$\ref{variability}),  we formed the median  average of these three highest quality spectra, i.e. the 3 repeats with corresponding control star observations.
The epoch-to-epoch variation is added to the calculated uncertainty at each wavelength, though in most cases this is less than the original Nod-Nod uncertainty. The spectral distribution of HD69830  from the visible to 100 $\mu$m is given in  Figure~\ref{HD69830SED}. The spectrum of the excess after subtraction of  the stellar photosphere is discussed in $\S$\ref{results}.  

\subsubsection{IRS Peak-Up Data}

While IRS serves primarily as a spectrograph, it also provides well-calibrated images  as it centroids its target onto the slit. As part of this peak-up procedure, we obtained IRS imaging data in the red filter (18.5-26.0 $\mu$m)   whenever HD69830 or HD68146 was observed in either the IRS High or Low  resolution modes (Table~\ref{ObsLog}). With each IRS spectra taking 6 snapshots,  a total of 84 images were obtained of HD69380 and 60 images of HD68146. We calculate the fluxes in each image  using aperture photometry  within a 4-pixel radius aperture  relative to a 8-to-14 pixel sky annulus,   using a conversion factor of 680 electron/sec per mJy (i.e. standard photometry as described in the IRS data handbook). Centroiding onto the peak of the flux in each image  moves the target location by less than a pixel with negligible effect on the net fluxes. Two frames with cosmic ray hits near the target were removed  (one from ``Repeat 3'' on HD69830 and one from ``Repeat 2'' on HD68146). Lastly, the individual frames were averaged into daily values, given as ``raw $F_{\nu}$'' in Table~\ref{IRSPeakup}.
HD68146 was observed whenever  HiRes observations were made allowing for careful photometric calibration monitoring at those epochs.  

The measured brightness of HD69830 at 22 $\mu$m is 270.4$\pm$3.1 mJy, with a dispersion of 1.4\% between April 2004 and January 2009.
For epochs with observations of both HD69830 and HD68146, the flux measurements of the stars have dispersions of 1.2 and 1.1\%, respectively.
Although these levels of variability are higher than Hipparcos limits on their visible light dispersion  \citep[$<$0.60\% and $<$0.65\% respectively;][]{perryman97}, the similarity between the two stars' dispersions suggests an origin  that is instrumental, not physical.
As further evidence of instrumental effects, the variations between epochs are correlated between the two stars.
The strongest deviations come during ``Repeat 3b'' (2008 Apr 30)  when {\it both} stars are about 1\% higher than at other times.
Hence, for measurements of HD69830 with contemporary observations of HD68146, some additional photometric calibration is possible.
Normalizing the HD69830 data using the HD68146 observations where available leads to the corrected values for HD69830 in Table~\ref{IRSPeakup} and Figure~\ref{IRSPeakupFig}.  The dispersion in the renormalized data for HD69830 (total of star $+$ disk) is just 0.9\% (1 $\sigma$) which should be taken as an upper limit on the mid-IR variability of the system as a whole. We discuss the possible variability of the disk emission in Section \ref{outburstlimits}.

\subsubsection{IRS High Resolution Spectra}

Three different epochs of HiRes data of both HD69830 and HD68416 were collected on 2007 December 20, 2008 December 4, and 2008 December 13. 
The first two epochs each included two observations at two Nod positions with the final epoch including one observation of two Nod positions for a total of ten spectra of each star. An observation on  2008  April 30 had downlink problems and is not considered further. Each observation also included an exposure offset from the star to be used for  background and bad pixel removal. All images were run through {\it IRSclean}\footnote{http://irsa.ipac.caltech.edu/data/SPITZER/docs/dataanalysistools/tools/irsclean/irscleanmanual} to flag and repair bad pixels using the bad pixel masks provided by the Spitzer Science Center (SSC) for each observational campaign, as well as bad pixels flagged by hand using {\it IRSclean}. After bad pixel repair, the background images were subtracted from their corresponding target images. The 1-D spectra were then extracted from the reduced images using  {\it Spice}\footnote{http://irsa.ipac.caltech.edu/data/SPITZER/docs/dataanalysistools/tools/spice/spiceusersguide} and its default settings.

Prior to stitching together the ten orders within the Long and   Short wavelength  modules, the 1-D spectra are put through {\it IRSfringe} to remove the low level of interference fringes inherent to the data.  At this point, there were still spikes in the spectra due to missed bad pixels and noise at the red   end of each module.  To  remove the remaining errant pixels in the 1-D spectra, we ran each module through a three pixel 3-$\sigma$ clipped mean. We stitched orders together using the data at overlapping wavelengths in each lower wavelength order to  attach onto its neighboring longer wavelength order. 
At a few wavelengths  in the overlap regions  we interpolated values between neighboring orders to provide a smooth transition;  the uncertainties assigned to these wavelengths were increased by 50\%, a conservative value based on examination of other overlap regions.  The same procedure was used to attach the Long and   Short modules, 
resulting in  ten complete spectra ranging from 9.86 to 37.2 $\micron$.  The final  spectrum was produced by first normalizing each spectrum to have the same flux at 15 $\mu$m, a region of the spectrum relatively free of residual noise features, and then taking a 3-$\sigma$ clipped average of the ten sub-spectra. Uncertainties were obtained by evaluating the standard deviation of the mean for the spectra included in the average at each wavelength. As a final step, the entire HiRes spectrum (star+disk) was adjusted by 3.5\% to bring it into agreement with the LoRes spectrum, an  amount which is within the quoted 5\% calibration uncertainties of the HiRes spectrophotometry \citep{Houck04} and with our contemporaneous HiRes observations of HD68146.

\subsection{Ground Based Observations}\label{ground} 

\subsubsection{Gemini/Michelle}

Observations of HD69830 were made using the Michelle instrument on the Gemini-North telescope. Images were obtained on two nights (7-8 Mar 2007) at 11.2 and 18.5 $\mu$m. Careful flat-fielding, registration and co-addition of individual frames resulted in images that proved to be unresolved compared to the reference star, HD61935. Some frames with significant trailing blur in the direction of motion of the chopping secondary were rejected. The flux densities detected in  2\arcsec\ diameter apertures (Table~\ref{phottable}) are consistent with the total star+disk flux density measured in the larger beams with Spitzer, confirming  the compact nature of the disk. The azimuthally averaged 1-dimensional scans of HD69830 and HD61935 at 18.5 $\mu$m are shown in Figure~\ref{Michelle} and are consistent with unresolved sources with a Full Width at Half Maximum (FWHM) of 0.3\arcsec. The limit to the size of HD69830's emitting region at 11.2 $\mu$m is 4 AU  diameter,  which is consistent with the inferred temperature and location for the dust discussed in the models below.  Similar results were reported by \citet{smith08}. As discussed below, mid-IR observations with the MIDI Instrument on the VLT-Interferometer indicate that the emission is resolved at the 0.5-1 AU scale \citep{smith09}.

\subsubsection{Keck Interferometer}\label{kidata}

While HD69830 is  too faint for observations with the nulling mode of the Keck Interferometer   (KI) at 10 $\mu$m \citep{colavita09}, we did obtain  visibility measurements  on KI's 85 m baseline in the $K$ band on 2006 Nov 11. Three integrations were made  in the medium resolution mode (R $\sim$ 200).  HD68146 and HD71155 were used as calibrators, chosen to match the target near-infrared brightness and spatial location, and
the wide-band and spectroscopic observations were processed with the standard settings, including a correction for flux bias.\footnote{http://nexsci.caltech.edu/software/KISupport/dataMemos/fluxbias.pdf}
Although HD71155 has a mid-infrared excess \citep{rieke2005}, its spectral energy distribution shows no evidence for a near-infrared excess and the uncertainty in its stellar size is not the dominant term in the
final uncertainty for the measured visibility of HD 69830.
The calibrated average visibility squared for HD69830 is 1.01 $\pm$ 0.05. For a stellar radius of 0.87 $\pm$ 0.04 R$_{\odot}$, the visibility squared for the stellar photosphere at $K$ band should be 0.97 $\pm$ 0.01.  Thus our measurements are consistent with a completely unresolved source and no excess flux.  The 1 $\sigma$ uncertainty on incoherent (i.e.\ distributed) flux within the 50 mas FWHM beam (0.63 AU) is 3\% of the stellar flux. Observations in the $L$ band were taken at KI on 26 Feb 2010.  The same calibrators were used as for the $K$ band observations and 5 calibrated scans on HD69830 were obtained. The data were processed as for $K$ band, except that no flux bias was applied.  The central wavelength is 3.91 $\micron$ and the source was unresolved with an average visibility squared of 0.97 $\pm$ 0.05.  The 1 $\sigma$ uncertainty on incoherent flux within the 90 mas FWHM (1.13 AU) is 3\% of the stellar flux.

\section{Results}\label{results}

\subsection{The Excess Toward HD69830}

As in earlier work \citep{beichman05irs}, we fitted a Kurucz model photosphere \citep{kurucz1992,castelli03} to photometric data from the visible and near-IR using Hipparcos and 2MASS data. Because the 2MASS data are of low quality due to the brightness of the star, we included data from the two short IRAC bands, 3.5 and 4.5 $\mu$m in the fitting procedure. The overall fit is of excellent quality with a reduced $\chi^2$=1.29 (Figure~\ref{HD69830SED}). The shortest wavelength module (SL2; wavelengths $<$ 7.5 $\mu$m) shows no evidence for features or any excess flux $\gapp$10 mJy or, equivalently, a fractional excess at roughly the 0.5\% level (1$\sigma$)  compared to the photosphere.  This level of excess is insignificant in terms of disk properties due to the brightness of the  photosphere at these wavelengths. 
The mid-infrared and longer wavelength photometry are summarized in Table~\ref{phottable} with amounts of excess given relative to this model. Most striking is the lack of evidence for any excess in the two longer IRAC bands, 5.7 and 7.87 $\mu$m. The onset of the excess seen in IRS modules (below) is quite sharp and shows no evidence for emission shortward of  $\sim$8 $\mu$m. The photometric data provide a  justification for  pinning the IRS spectra to the photosphere at the short end of SL1 (as described below). The lack of excess shortward of $\sim$8 $\mu$m (in both the IRAC and IRS data) supports the contention that amorphous carbon is not present in this source \citep{lisse07}. 

The broad outline  of the excess toward HD69830 is revealed in the low resolution spectral energy distribution (Figure~\ref{HD69830SED}) which shows an excess between 8 and 35 $\mu$m that by 70 $\mu$m falls off to levels below detectability, even  with the new, higher SNR MIPS data (\S\ref{mips}). This
confirms the earlier lack of long wavelength excess \citep{bryden06}.  As argued in \citet{beichman05irs}, these characteristics require the existence of small grains located within 1 AU of the star.  The lack of significant 70 $\mu$m excess means both that the emitting grains detected at 8-35 $\mu$m  must have sharply declining emissivity at longer wavelengths (and hence be much smaller than $a\sim 70/2\pi \sim 10 \, \mu\rm{m})$  
and that there is no large reservoir of larger, colder grains at greater distances from the star.  The plethora of small grain features in the low and high resolution spectra  confirm these conclusions and are discussed below.

\subsection{The Low Resolution Spectrum}\label{lores2}

The averaged LoRes spectrum from three epochs (Dec 2007 to Dec 2008)  is shown in Figure~\ref{HD69830LoRes}a after subtraction of the stellar photosphere.  The fractional excess relative to the photosphere is shown in Figure~\ref{HD69830LoRes}b.  The spectrum  is in good agreement with the calibration of the peak-up image with an average flux across the 18-26 $\mu$m passband of the peak-up array of 83 mJy, compared with the 88$\pm$8 mJy measured in the peak-up images.  The average 7 to 35 $\mu$m spectrum  shows qualitatively the same features as in \citet{beichman05irs} with numerous features from small grains evident throughout the spectrum. Improvements from the additional observing time include higher signal to noise ratios at longer wavelengths and better control of systematics around 8 and 14 $\mu$m.  We use this improved spectrum for mineralogical analysis in $\S$\ref{model} below. 

We put a limit on any temporal variations across the 1 year observing sequence for the best LoRes data by noting that the  uncertainty in the average excess, $\sigma F_{dust}/ F_{dust}$, averaged over the ranges with significant excess, 9.5-11.5 \micron\ and 16-32 \micron, is 7\% and 5\% per spectral element, respectively.  These variations  are only slightly larger than the variation seen in the spectra of HD68146 which showed no fractional excess at the level of $2-3$\% on three separate occasions across the whole 7 to 35 \micron\ band. Figure~\ref{HD69830LoRes}b  shows that the differences in the fractional excess  from each spectrum relative to the average is $\sim$1\% in regions of good SNR and away from a few bad pixels. The average LoRes spectrum is available in electronic form (Table~\ref{IRSLRSdata}).

\subsection{The High Resolution Spectrum}

To facilitate the search for narrow spectral features in the HiRes spectrum, we  subtracted a  spline curve fitted to  a running  median-smoothed (25 point) version of the photosphere-subtracted spectrum. We  applied a 5 point smoothing (Hamming) filter with  weights (0.04, 0.24,0.45, 0.24,0.04), corresponding roughly to the HiRes instrumental response, for further rejection of bad pixels and residual fringing. Close-ups of the filtered, star-subtracted spectrum are   shown in Figure~\ref{FlattenHiresPanels}. The local maxima in this spectrum do not stand out above  the noise to a  significant degree, nor do they line up with any  prominent known spectral features. In most cases  local maxima at $\sim3\sigma$ can be traced back to  artifacts in the raw data or to complex structure in the small-grain emission. Upper limits on the fluxes at interesting line locations are given in Table~\ref{HiResLines}. Simulated spectral lines at the level of our upper limits are shown (Figure~\ref{FlattenHiresPanels}).

\section{Discussion}\label{discuss}

\subsection{Model for Emission}\label{model}

Our existing models \citep{beichman05irs, lisse07} required the presence of hot, small crystalline grains in the region of 1 AU with no evidence for material on more distant orbits. The new observations confirm and extend these results.

\subsubsection{Re-Examination of HD69830 Dust Mineralogy}

The physical model  used to characterize the mineralogy of the HD 69830 dust was developed for Spitzer observations of the Tempel 1/Deep Impact data \citep{lisse06}.  In our case, optically thin emission originates in  material  located in a narrow annulus around the star. A wide variety of minerals and ices are successively added into the model until the  $\chi^2$ of the fit stops changing by a significant amount. Additional variables include the size distribution of the dust grains and distance to the host star. Our model of the 2004 data  \citep{lisse07} revealed a population of small, super-thermal grains at about 425 K composed of silicates (both olivine and pyroxene, but mainly forsterite, fayalite, and ferrosilite), a smattering of sulfides and carbonates, and surprisingly some water ice. 

Using the higher SNR LoRes spectra (\S\ref{lores2}), we now reevaluate the mineralogy of the HD69830 system. Figure~\ref{fit} shows the best fit model broken down into its various species components while Table~\ref{caseytable} lists the relative abundances of species identified in the LoRes spectrum, as well as significant non-detections. The final reduced $\chi^2$ for the model fit to the data is 1.01. For the most part, our results are the same as in our previous work, with values and upper limits improved due to the much higher quality and SNR of the 2007 data. Each of the three 2007 spectra included in Figure~\ref{HD69830LoRes}  had much longer  integration times than did the 2004 data, e.g. 70 and 480 seconds  in the SL1 and LL1 modules in 2007 vs. 12 seconds and 28 seconds in 2004. Differences from the earlier analysis are flagged in the table and  include the new (albeit weak) detection of phyllosilicate talc and iron oxide and the failure to confirm a number of minor carbonates and pyroxene species. The one important difference in the new analysis is the lack of convincing evidence for the broad feature due to water ice (11-15 $\mu$m), a spectral region now understood to have serious problems with the tear-drop artifact in the data (\S\ref{lores}). The new spectra show no obvious emission from water ice, which is significantly different than the model findings reported in \citet{lisse07}. The newer data, when compared to the old, show shifts in the 10-35 $\mu$m region where water ice emission is most pronounced. As the other mineralogical markers seem to show that the dust in the system has not changed appreciably, we believe the difference is due to the higher SNR of the new data, and to a better, more complete IRS data calibration. While previously we had reported water ice at only moderate abundances in HD69830, the change does concern us for the purported detections of water ice in other systems we have studied to date.

Figure~\ref{dustsize} shows the contributions of various particle sizes to the overall emission spectrum along with various power-laws. The steepest, $a^{-3.9}$, power-law provides the best fit to the IRS spectrum. The distribution of grains sizes is typically assumed to follow a power-law distribution $dn/da \propto a^{-3.5}$, as expected from a simple model for a steady-state collisional cascade \citep{dohnanyi}. However, a greater concentration of small dust can be produced if the strength of the particles against collisional disruption depends on grain size, with smaller particles being more resistant to destruction than larger ones \citep{obrien2003,wyatt2011,belyaev2010}. Observationally, a steeper power-law appears to be required to explain the combined IR and visible light emission from the resolved disks orbiting HD207129 ($a^{-3.9}$) and HD92945 ($a^{-3.7}$) \citep{krist2010, golimowski2011}. Our model for HD69830's disk emission suggests a comparably steep power law.  Consistent with our previous work we obtain a dust mass of 3$\times10^{20}$ g and surface area of 1.2$\times10^{24}$ cm$^2$ in 0.1-10 $\mu$m grains and, by extrapolation up to 10 m planetesimals with a $a^{-3.9}$ power law, a total mass of solid material 2$\times10^{21}$ g \citep{lisse07}.

Another explanation for the particle size distribution comes from the competition between gravitational and radiation forces (Poynting-Robertson drag and blow-out by radiation pressure)  which controls the small grain lifetime.  The parameter $\beta= F_{rad}/F_{grav} $  depends on physical properties of the dust grains and the stellar radiation field  \citep{wyatt1999,sheret04}. The value of $\beta$ as a function of particle size is shown in  Figure~\ref{betafig} for astronomical silicate \citep{draine84} in the radiation field of  HD69830. Only a narrow size range around $\sim$ 0.4 $\micron$ diameter  attains $\beta>$0.5, so that a turnover (but not a complete truncation) in the size distribution would be expected at sizes just below 1 $\micron$ as observed; i.e., the derived distribution is entirely consistent with that expected for dust that is being produced in steady state from a reservoir of larger objects. For different compositions, in particular those with grains of lower densities, $\beta$ would be higher than that in Figure~\ref{betafig} and a broader range of particle sizes would be removed.

Overall, the new data, particularly given the lack of a definite detection of H$_2$O ice, leads to a straightforward conclusion that the HD69830 debris is dominated by small grains $<$1 $\mu$m composed of material similar to C-class asteroids which formed in drier, interior portions of solar nebula. As suggested by Figure 10 in \citet{currie2011}, the silicates in the HD69830 disk are enriched in olivines relative to pyroxenes, suggestive of highly processed material in a few Gyr-old system and substantially different from the more primitive material seen in the comet Hale-Bopp or the young stars HD 98800 and $\eta$ Crv. 

\subsubsection{Location of the Emitting Material}

Mid-IR interferometric results from the MIDI instrument on the VLTI \citep{smith09} resolved the HD 69830 disk and, in combination with stability constraints from the known planets (Lovis et al.\ 2006), suggested two possible locations for the emitting material: in the middle of the planetary system at 0.5 AU and beyond the outer planet outside 1 AU, with the closer location being marginally preferred by the VLTI data. We reexamined the possibility of the grains being at $\sim$0.5 AU, inside the planetary system, rather than our preferred solution of grains located at 1 AU (near the 2:1 resonance with the outermost planet). The equilibrium temperature of small grains that are inefficient emitters is given by \citet{backman93}:

\begin{equation}
T_{gr}= 468 \,\, \left( \frac{L_{\star}}{L_{\odot}} \right)^{1/5} \, 
 \left(\frac{AU}{R}\right)^{2/5} \lambda_0^{-1/5}\, K
\end{equation}

\noindent where $\lambda_0 = \zeta a \sim 0.1\, \mu$m is the critical wavelength for $a$=0.1 $\mu$m grains and $\zeta\sim 1$ is appropriate for moderately absorbing grains \citep{backman93}. Then $T_{gr}=670 \, R_{AU}^{-2/5}\, K$ for HD 69830's $L_{\star}=0.6 L_\odot$ \citep{lovis06}. Putting the grains at 0.5 AU would raise their temperature to over 800 K which is inconsistent with the lack of emission shortward of 7.5 $\mu$m. Alternatively, larger grains at 0.5 AU would have lower temperatures ($\sim 350$ K), but these large particle sizes are totally inconsistent with the sharp crystalline features seen in the spectrum which demand grains roughly $\sim$0.1-1.0 $\mu$m in radius. The detailed modeling reaffirms this conclusion and strongly supports the 1 AU location of the debris, exterior to the 3 planets and not between them. The SED analysis favors the alternative interpretation of the VLTI results that places the bulk of the material at the more distant location from the star.

The data from the Keck Interferometer (KI,\S\ref{kidata}) yield upper limits to the presence of material emitting at 2 and 3 $\mu$m of $F_d/F_* \sim3$\% ($1\sigma$) of the photospheric levels. Using a standard relationship \citep{beichman06irs}, this flux ratio limit can be converted to a limit on the amount of material (in units of $L_d/L_* $) emitting at the temperature, $T_d$, at which the Planck emission is at its peak:

\begin{equation}
\frac{L_d}{L_*} = \frac{F_d}{F_*} \frac{e^{x_d}-1}{x_d} \left( \frac{T_d }{T_*}\right)^3 
\end{equation}

\noindent where $x_d=h\nu/(kT_d)=3.9$ at the peak of the Planck curve. For dust emitting at 2.2 (3.1) $\mu$m, with peak temperatures of 1700 (1200) K, the Keck Interferometer measurements yield 1 $\sigma$ limits to $L_d/L_* $ of 0.01 (0.004). These temperatures are reached at distances from the star of $\sim$ 0.02 (0.04) AU for a blackbody or 0.1 (0.2) AU for small grains. Unfortunately, these limits do not particularly constrain the amount of material in the innermost regions of the HD 69830 disk, compared with the ten-fold smaller amounts of hot dust seen toward the A and F stars targeted by other interferometers \citep{absil2008,akeson09}. These nearby hot stars are 3-5 magnitudes brighter than HD 69830 and were observed with interferometers with higher $V^2$ precision than KI. The KI 2 and 3 $\mu$m results are also consistent with the shortest wavelength VLTI results (8 and 9 $\mu$m) of \citep{smith09} which showed the source to be unresolved.

\subsubsection{Limits to Gas Species from HiRes Data}

Table~\ref{HiResLines} lists spectral features used to search for gas phase constituents in the HD 69830 disk where we have included typical atomic and molecular species seen in younger disks such as TW Hya and other T Tauri stars \citep{chen2006,najita2010,pascucci2007}. In no case do we find convincing evidence for the presence of any gas species. Using the model described in \citet{salyk08} we have turned our measurements into upper limits for a few key species using an optically thin model that depends solely on the assumed gas temperature (Figure \ref{gaslimit}). For a gas temperature of 400 K, appropriate to gas in equilibrium with dust grains at 1 AU, we find an upper limit to the amount of water vapor present of 2$\times10^{17}$~g. The minimum dust mass derived by \citet{lisse07} and consistent with the new data reported here is $3\times10^{20}$ g (for grain radii 0.1 to 10 $\mu$m), 1500 times the limit to the amount of water vapor. The HD 69830 debris field is at present very dry. To assess the meaning of this upper limit in terms of a possible parent body, we must examine the lifetime of any water vapor against photolysis by the combination of the radiation from HD 69830 and the interstellar radiation field. 

We first consider the photolysis rate of optically thin water gas in the interstellar radiation field:
$dN/dt = -J N$,
where $N$ is the water abundance in cm$^{-3}$ and $J = 5.9 \times 10^{-10}$~s$^{-1}$ is the experimentally measured rate coefficient \citep{millar97}. 
The half-life against photodestruction by interstellar photons is then 
$\left( \ln 2 \right ) / J =$ 37.2 yr.

Next we consider photodissociation from the star's UV flux.
For cometary material located at 1 AU from the Sun, the time scale for dissociating H$_2$O, $\tau(H_2 O\rightarrow $OH + H), is $ 8 \times 10^4$ sec or just under 1 day \citep{Combi93}. The stellar photon field at 1 AU from the cooler HD 69830 is approximately 5 times weaker shortward of the 0.24 $\mu$m photodissociation limit for H$_2$O than the Sun's. In calculating this ratio we used the ratio of a 5750 K Kurucz model (representing the Sun) to an average of 5250 K and 5500 K models, which is consistent with HD 69830's IUE measurement of F$_\nu$(0.24 $\mu$m)=42 mJy \citep{IUE}. By this reckoning, the survival time for water will be approximately $4 \times 10^5$ sec or $\sim$4.6 days.

Thus, while a comet explosion, e.g. Comet 17P/Holmes in 2007 \citep{Reach2010}, or collision could produce IR-emitting dust plus H$_2$O gas, the water lifetime will be so short that just a few weeks after the collision all traces of the water would have vanished. Models of the solar nebula suggest that the ice/rock ratio of icy bodies is around 1:1 \citep{lodders03, sally09ice}, so we assume that such a comet explosion would produce equal amounts of water gas and IR-emitting dust - $3 \times 10^{20}$ g of each for HD 69830.
Although this amount of water gas is well above our detection threshold of $\sim$$2 \times 10^{18}$~g and would have been observable, the time scale is so short that catching the water emission is unlikely.

The lack of detection of H$_2$O also constrains scenarios in which the dust and gas are replenished in a steady state, with the equilibrium gas mass balanced between the mass input and photodissociative loss. Assuming the  parent body composition is equal parts ice and dust, and that 100\% of the 
ice ends up as water vapor, the inferred dust mass loss rates of between  10$^{15}$ g/day (Wyatt et al.\ 2007a) and 10$^{18}$ g/day (Wyatt et al.\ 2010) also  apply to the water vapor production rate. The 4.6 day lifetime of water 
vapor then predicts steady state levels of $5\times10^{15}-5\times10^{18}$ g. The upper limit implies that mass loss rates as high as that invoked in Wyatt  et al.\ (2010), in which the dust we see is being blown out via radiation  pressure, are ruled out unless the parent body had a particularly low water content.

OH is thought to arise from dissociation of H$_2$O \citep{carr2011}, so  the lack of OH in our spectrum is not surprising given the lack of H$_2$O. Similarly, species such as H$_2$C$_2$ and HCN are seen in T Tauri stars at a few percent to $\sim$50\% of the level of H$_2$O emission \citep{carr2011}, so that the lack of these species is also consistent with the non-detection of water and OH. While we cannot rule out trace amounts of water ice and gas in the HD 69830 debris or a one-time burst of cometary material from which the generated gas species have been rapidly destroyed, the overall debris composition and the present day lack of gas is consistent with an origin in asteroidal not cometary parent bodies.

\subsection{A Search for Temporal Variations}\label{variability}

In addition to the numerous spectral features which we can use to determine the composition of its dust, HD 69830's disk is also unique in the proximity of the dust to the star and to known planets which permits a search for temporal evolution in the disk emission on dynamical timescales. The periods of the planets at 0.078, 0.19 and 0.63 AU are 8.7, 31.6 and 197 days, respectively \citep{lovis06}. While most debris disks are thought to be evolving in steady state due to the slow grinding down of planetesimal belts \citep[e.g.][]{dermott02, dominik03,wyatt07b}, there is a growing number of disks that exhibit evidence that the IR emission must be transient \citep{telesco05, su05, lisse09}. HD 69830 may be one such transient disk \citep{wyatt07hot}. Of the possibilities for the origin of this ephemeral dust, several predict variability on orbital timescales: (i) If the dust originates in planetesimals that are trapped in resonance with the planet HD 69830d at 0.63 AU then its distribution would be clumpy, and those clumps would orbit the star with the planet \citep{wyatt02}. If the planet's orbit is not exactly circular, then those clumps would also be on eccentric orbits \citep{quillen02, kuchner03}, and so the dust temperature and total emission would vary on the orbital timescale of the planet ($\sim$200 days). Such variations would also affect the collisional/sublimation properties of the feeding material which could be evident in changes in spectral features on the same timescale. (ii) Similar temperature and emission variability would be predicted if the dust originates in a clump of material on an eccentric orbit (e.g., if this is the recent disintegration of a comet scattered in from a more distant planetesimal belt), except that the orbital timescale could be much longer. (iii) Since the majority of the dust is small enough to be blown out by radiation pressure as soon as it is created, then if the small dust population in this planetesimal belt is dominated by individual collisions \citep[e.g.,][]{song05} there would be a significant stochastic element to the emission, which could originate in the destruction of planetesimals as small as 30 km \citep{lisse07}.

\subsubsection{Limits on Outbursts}\label{outburstlimits}

As \citet{dermott02} have noted ``extrasolar planetary systems, even very old systems, that contain belts of asteroids and comets in the rubble pile state could occasionally and unexpectedly flare into infrared visibility due to catastrophic disruption of these rubble piles.'' Such an event must be considered as the cause of the brightness of HD 69830 \citep[see also][]{Kenyon04}. However, there is no evidence for large scale changes in the disk brightness on a number of interesting timescales. First, over the 24 year period between IRAS and the latest Spitzer  data, from 1983 to 2007, there is no evidence for variation in the disk output: MIPS (24\micron), F$_\nu$(disk)=0.078$\pm$0.005 Jy, and IRS peak-up  (22\micron), F$_\nu$(disk)=0.088$\pm$0.008 Jy, compared with IRAS 25 $\mu$m, F$_\nu$(disk)=0.100$\pm$0.026 Jy (Table~\ref{phottable}). Ignoring the slight differences in wavelengths, the variation is insignificant ($16\pm28$ mJy) compared to an average disk value between 22 and 25 \micron\ of 84 mJy. There is no evidence for variability of the disk emission at the 33\% (1$\sigma$) level over 24 years where the large uncertainty is due to the low SNR of  the IRAS data.

IRS peak-up photometry spans only 4 years but provides information at the few percent level. As discussed above, the IRS peak-up data (22 $\mu$m) shows no significant change relative to the IRAS data (25 $\mu$m) over a 24 year timescale (Figure~\ref{IRSPeakupFig}a).  To investigate the possible variation of the excess emission with higher precision on the shorter Spitzer timescale, we examined the IRS peakup data corrected using contemporaneous HD68146 data (when available). After subtracting the photospheric value of 182.2 mJy for HD 69830 we find that the disk emission is consistent with a {\it constant} value of 88 mJy at 22 $\mu$m, with 2.6\% measurement dispersion. Including the uncertainty in the stellar contribution increases the uncertainty in the disk excess to 3.3\%, or 88$\pm$3 mJy (Figure~\ref{IRSPeakupFig}b). The $\chi^2$ value of the data compared against the assumption of constant disk emission is 9.4 which with 6 degrees of freedom has a non-significant P-value of 0.15. Finally, the IRS LoRes spectra show no difference in the dust emission at the 5-7\% level per spectral element over a 1 yr period (Figure~\ref{HD 69830LoRes}b).

The total emitting area in the HD 69830 dust cloud is estimated to be $1.2 \times 10^{20}$ m$^2$ in particles ranging in radius from 0.1 to 10 $\mu$m \citep{lisse07}. The lack of observable change in this emission implies that the surface emitting area has changed by less than $<3\times 10^{18}$ m$^2$ (4 years) and $<40\times 10^{18}$ m$^2$ (24 years). Since the total disk mass corresponds to an object with a roughly 30 km radius, we can set limits on the complete destruction of an object of radius equal to 10 (20 km) and converted into small grains on time scales of 4 (24) years. The destruction of such an object corresponds to the material responsible for the zodiacal dust bands in our solar system \citep{grogan01}. A catastrophic event of this sort could be expected to occur once per 1-5 million years in the solar system \citep{dermott02, sykes86}. In the dynamically active planetary system of HD 69830 such an event might occur tens to hundreds of times more frequently. While the probability of catching such a rare event is small, considering that HD 69830 is itself a rarity with only $<1$\% of stars in various Spitzer surveys showing bright 10 $\mu$m excess emission \citep{beichman06irs, lawler09}, we argue here and in $\S$\ref{nature} that this explanation may be the most probable for this star.

\citet{heng2011} argues that a steady state grinding down of material is a possible explanation for the HD 69830 excess, but only if the star has an age of $\sim$1 Gyr. In this scenario, the disk would retain more of its primordial brightness \citep{wyatt02}. While age estimates for HD 69830 range from $<$1 to 5 Gyr \citep{beichman05irs,wyatt07hot}, older ages seem much more probable. One new piece of information comes from a ``probable" detection of a stellar rotation period for HD 69830 of 35$\pm$1 day \citep{simpson2010}. This period is consistent with expectations for a K0V star with the age of the Sun or greater \citep{barnes2010}. Chromospheric and X-ray data similarly argue for ages $>>$ 1 Gyr. Measured values of log(R$^\prime_{HK})=-4.98$ and log(L$_X$/L$_{Bol}$)=-6 \citep{canto2011} both suggest ages of 4-5 Gyr \citep{maldanado2011, mamajek2008}. At such an advanced age, \citet{heng2011} agrees that a steady state model is hard to defend and that a transient origin for the HD 69830's excess is more likely.

Another source of an outburst could be the passage of a comet into the inner reaches of the planetary solar system \citep{beichman05irs}. The brightening of the solar system's zodiacal cloud due to incursion of comets has been examined quantitatively by \citet{napier01}. A major argument against this mechanism is the spectral character of the dust emission which is far more asteroidal than cometary in nature \citep[$\S$\ref{model}]{lisse07}  and the lack of any gas steady state emission. A weaker argument against a cometary origin is the fact that this system has no detectable dust associated with a distant Kuiper Belt. As discussed in Beichman et al. (2006), the 70 \micron\ flux density limit corresponds to $L_d/L_*\sim 5 \times 10^{-6}\,  (3\sigma)$ for 50 K dust. This level  is only  $\sim$5 times greater than the nominal range predicted for the  Kuiper Belt dust in our own solar system \citep{stern96}. The fact that the reservoir for comets is not much larger or more dynamically active than in our solar system argues that large scale cometary incursions must be rare.

\subsubsection{Limits on Orbital Effects}

One possible origin for temporal variability is that a fraction of the disk material could be differentially heated as it orbits the star on an eccentric orbit. In analogy with the zodiacal dust bands in our own solar system, most of the debris material 
will be spread uniformly along the orbit since the dynamical timescale for this spreading is just a few hundred years within an AU of the star 
\citep{grogan01}. Material in an eccentric orbit would be hotter close to the star and cooler far away from it. This effect is clearly seen in resolved maps of the Fomalhaut disk \citep{stapelfeldt04, marsh05} 
where the SE ansa is hotter than the NW one. But this steady-state spatial effect is not resolvable in any data presently available. Nor is temporal data available for systems on such long period orbits. However, if there were a clump of material, located, say, at a resonant point with respect to the various planets in the system, then the periodic passage of this material closer to and further away from the star would change the temperature and resultant emission from the dust.

For specificity, consider the possibility that the $\sim$88 mJy of excess detected in the IRS peak-up images at 22 $\mu$m consists of a constant component of 80 mJy from uniformly distributed dust at a semi-major axis of 1 AU plus another 10\% in a clump that moves in an eccentric orbit with semi-major axis of 1 AU and an eccentricity, $\epsilon$. A simple model that follows the clump through its orbit yields a time-varying amount of emission we could compare with the data in Figure~\ref{IRSPeakupFig}. The model puts the dust clump in a 2:1 resonance with the outermost planet seen \citep{lovis06} at a semi-major axis of 1 AU, a corresponding period of approximately 400 days and eccentricities of 0.1, 0.3, 0.6 and 0.9. The small grain temperature follows the radial power-law derived above, $T_{gr}=670 \, R_{AU}^{-2/5}\, K$. Following the clumped grains through multiple orbits yields the orbital location, temperature and total disk emission which can be compared to the Spitzer observations (Figure~\ref{PlanckOrbit}). The IRAS value is not constraining. The maximum variation in the observed excess for the different eccentricities ranges from 0.6\% ($e$=0.1) to 24\% ($e$=0.9). Of course, the orbital location of any clump (taken here to be a point source) is unknown, so to assess the importance of eccentricity and clumpiness requires an average over all possible clump locations. Figure~\ref{PlanckOrbitFluxchi2} shows the result of a Monte Carlo simulation comparing the model of variable emission with the observations as a function of eccentricity and the fraction of total material carried in an eccentric orbit, averaged over all starting locations. The contours show that only an extreme range of eccentricity and clump fraction can be ruled out. For simplicity the simulation assumes a point source clump; to the extent that the clump is spread out in orbital phase, the amount of variation associated with the inhomogeneity   will be reduced.

\citet{wyatt10} have argued that a highly eccentric swarm of planetesimals ($\epsilon>>0.9$) could be responsible for a long-lived debris disks around stars like HD 69830. Without such high eccentricities and in the absence of a rare collisional event producing a burst of small grains, collisional processes operating over the presumed few Gyr age of HD 69830 would result in dust levels much closer to what we now see in our own solar system \citep{wyatt07hot}. If a high eccentricity parent population proves to be the explanation for the HD 69830 disks, then the results here suggest that the parent bodies of the emitting dust must be spread quite uniformly along the orbital path to account for the lack of the variability monitored by Spitzer and IRAS. 

\subsection{Nature of the HD 69830 Disk \label{nature}}

The  statistics of Spitzer detections suggest that  the incidence of hot (12 $\micron$) excesses for {\it mature FGK stars},  $\eta_{hot}$, is  less than $\sim$1\%  \citep{beichman06irs, lawler09}. Old K stars with excesses like HD 69830's are  rare, consistent with theoretical expectations that debris disks fade away over a few Gyr, particularly at orbital distances of a few AU. What then explains the origin of the HD 69830 excess? We argue below that an impulsive event, i.e.\ a destructive collision of a modest sized (30 km radius) asteroid is the best explanation for the excess.

The lifetime of a debris disk as pronounced as HD 69830's is dominated by collisional effects.  The timescale (e-folding time) for dissipation of the  dust from a  cataclysmic collisional event at 1 AU is given by $t_{outburst}=\frac{T_{orbit}} {4\pi\tau_{eff}}\sim$ 400 yr  \citep{wyatt1999} for an effective dust optical depth of $\tau_{eff}=2\times10^{-4}$ \citep{beichman05irs}.  Since we do not know when or at what level the outburst began, we can only guess that the  small grains excess would be at an observable level for a duration of $5\sim$10$\times t_{outburst}\sim$ 2,000-4,000 yr.  A more realistic timescale for an observable small grain excess   includes the grinding down of larger debris into new dust and its eventual dissipation, $t_{decay}\sim$1 Myr \citep{grogan01}. If this sort of event occurs regularly  but with a duty cycle given by the observed rarity of hot excesses among mature field stars, i.e.\ $<1$\%, then the interval between observable events is $t_{decay}/\eta_{hot}>$ 100 Myr. Over the $\sim$4 Gyr lifetime of HD 69830 this might happen between $\sim$ 40 times. If each event represents  the destruction of a 30 km radius asteroid, the total mass lost would be $6\times 10^{-7}$  M$_\oplus$ which is only 0.3\% of the current mass of our solar system's asteroid belt \citep{kras2002}.  

This scenario has been modeled in detail for young stars by Kenyon and Bromley (2005) who find large spikes in grain production due to collisions between 10$\sim$100 km diameter asteroids, noting that {\it ``at 1 AU, occasional collisions between large objects produce observable changes in the mid-IR excess. Small, few percent, changes in the mid-IR excess should recur on timescales of tens to hundreds of years. Large changes are much less frequent and recur on thousand year or longer timescales}.'' For an  old system like HD 69830 with a greatly depleted disk, the time between significant events will be much longer and the fractional change in the excess much larger.

HD 69830 has a much simpler debris disk system than another nearby K star, $\epsilon$ Eridani (K2V) which has a single planet \citep{hatzes00},  a rich system of dust annuli ranging in radii from 3 -100 AU, and  a total dust mass of $\sim 5 \times 10^{-4}$ M$_\oplus$ \citep{backman09}. Because $\epsilon$ Eri is a much younger star, 0.85 Gyr \citep{difolco07}, than HD 69830, it is tempting to regard HD 69830 as the endpoint in the evolution of a debris disk system having an undetectable Kuiper Belt and revealing itself  only  occasionally through the outburst of a rare asteroidal breakup. Evolutionary models of debris disks around solar type stars suggest a 5-10$\times$ decrease in $L_{IR}/L_*$ \citep{wyatt07b} in the far-IR over the time period of 0.85 to 4 Gyr.   Even after  this level of decline, $\epsilon$ Eri's excess would still be observable suggesting it started with a more massive protoplanetary nebula than  HD 69830's.

\section{Conclusion}\label{conc}

The main-sequence, solar-type star HD 69830 has an unusually large amount of dusty debris orbiting close to the host star and
the three known planets in the system. In order to explore the dynamical interaction between the dust and planets, we have performed multi-epoch photometry and spectroscopy of the system over several orbits of the outer dust. We find no evidence for changes in the dust amount or composition, with an upper limit of 3.3\% (1 $\sigma$) on the variability of the dust emission over a 4 year period and $<$33\% (1 $\sigma$) over the 24 year span between IRAS and Spitzer. Higher SNR observations have allowed us to refine the mineralogical composition of the debris cloud, confirming our identification of HD 69830 as being rich in olivine and other silicates. The one difference between the results described here and our earlier work is the lack of a convincing presence of water ice for which we can now only set an upper limit. On the basis of our mineralogical assay, we identify the origin of the dust as coming from material similar to that found in C-type asteroids in our solar system. Finally, we have explored the spatial distribution of the emitting material. Our spectral modeling is consistent with material located around 1 AU, outside the three planets and perhaps in 1:2 or 3:2 resonances with the outermost planet. We argue that we are witnessing a relatively rare asteroid collision produced, perhaps, via interactions between the planets and an asteroid belt that survived the formation and migration of the planets \citep{kenyon2005,ailbert06}.

\section{Acknowledgments}

\acknowledgments {We are grateful to an anonymous referee for a careful reading of this manuscript that led to significant improvements. This publication makes use of services provided by the NASA Exoplanet Science Institute at the California Institute of Technology (NExScI) and data products from the NASA/NExScI Star \& Exoplanet Database (NStED), the Two-Micron All Sky Survey (2MASS), and the NASA/IPAC Infrared Science Archive (IRSA). Thanks to Ben Oppenheimer, Dimitar Sasselov and Dave Latham for their hospitality and support during extended visits to AMNH and CfA, respectively. Some of the research described in this publication was carried out at the Jet Propulsion Laboratory, California Institute of Technology, under a contract with the National Aeronautics and Space Administration. The Keck Interferometer is funded by the National Aeronautics and Space Administration as part of its Exoplanet Exploration program. The authors wish to recognize and acknowledge the very significant cultural role and reverence that the summit of Mauna Kea has always had within the indigenous Hawaiian community. We are most fortunate to have the opportunity to conduct observations from this mountain. Copyright 2011 by California Institute of Technology. Government sponsorship acknowledged.

\facility{{\it Facilities:} Spitzer, Gemini/Michelle, Keck Interferometer}

\appendix


\clearpage
\begin{figure}
\includegraphics[width=1.0\textwidth]{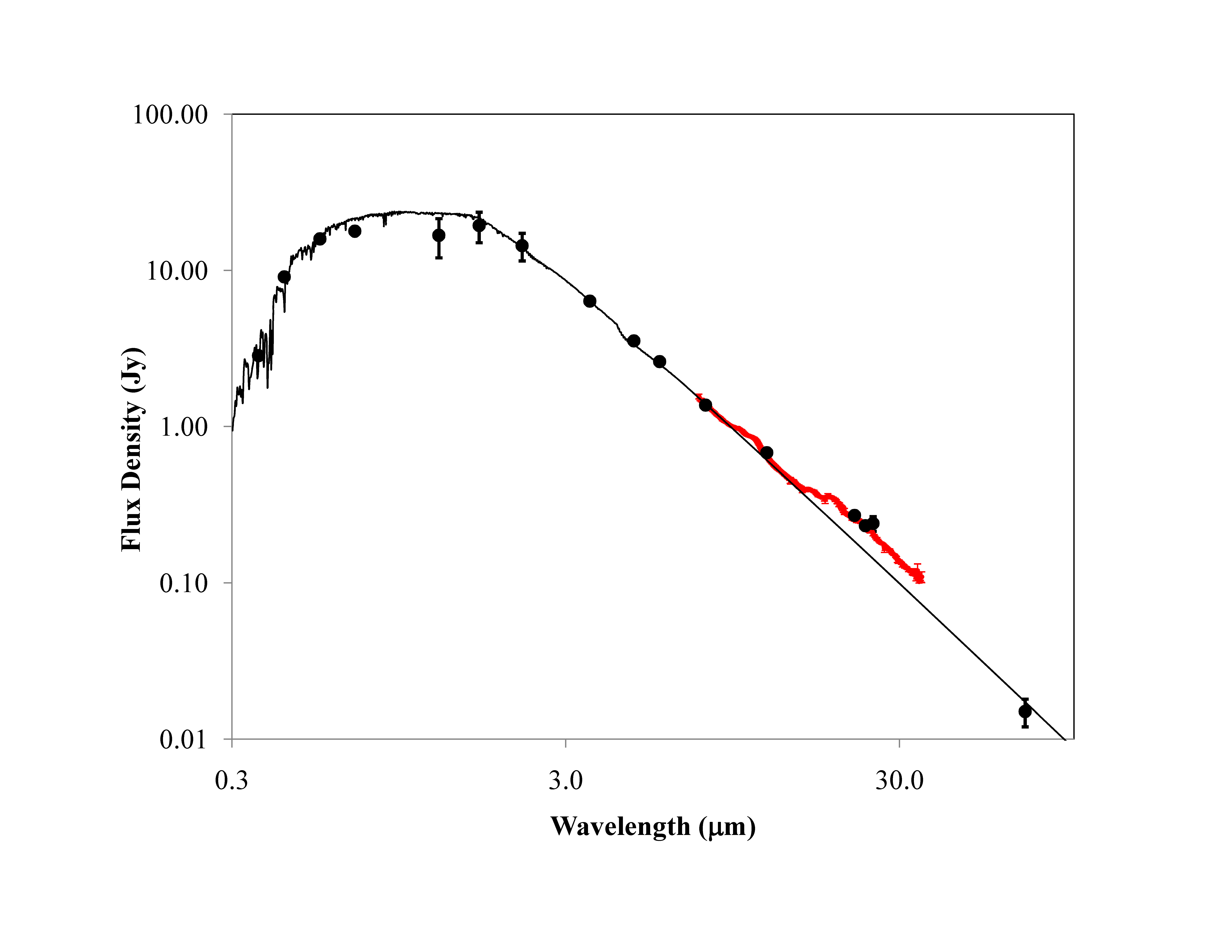}
\figcaption{{The full spectrum of HD 69830 includes a Kurucz photospheric
model fitted to short wavelength data ($<$ 5~$\mu$m) along with data from the three Spitzer instruments. The spectral energy distribution shows an excess at mid-IR wavelengths due to small grains within 1 AU of the star. The lack of an excess at 70 $\mu$m implies there is no significant reservoir of large, cold grains at larger distances from the star.}
\label{HD 69830SED}}
\end{figure}


\clearpage
\begin{figure}[h]
\begin{center}$
\begin{array}{c}
\includegraphics[width=3.5in]{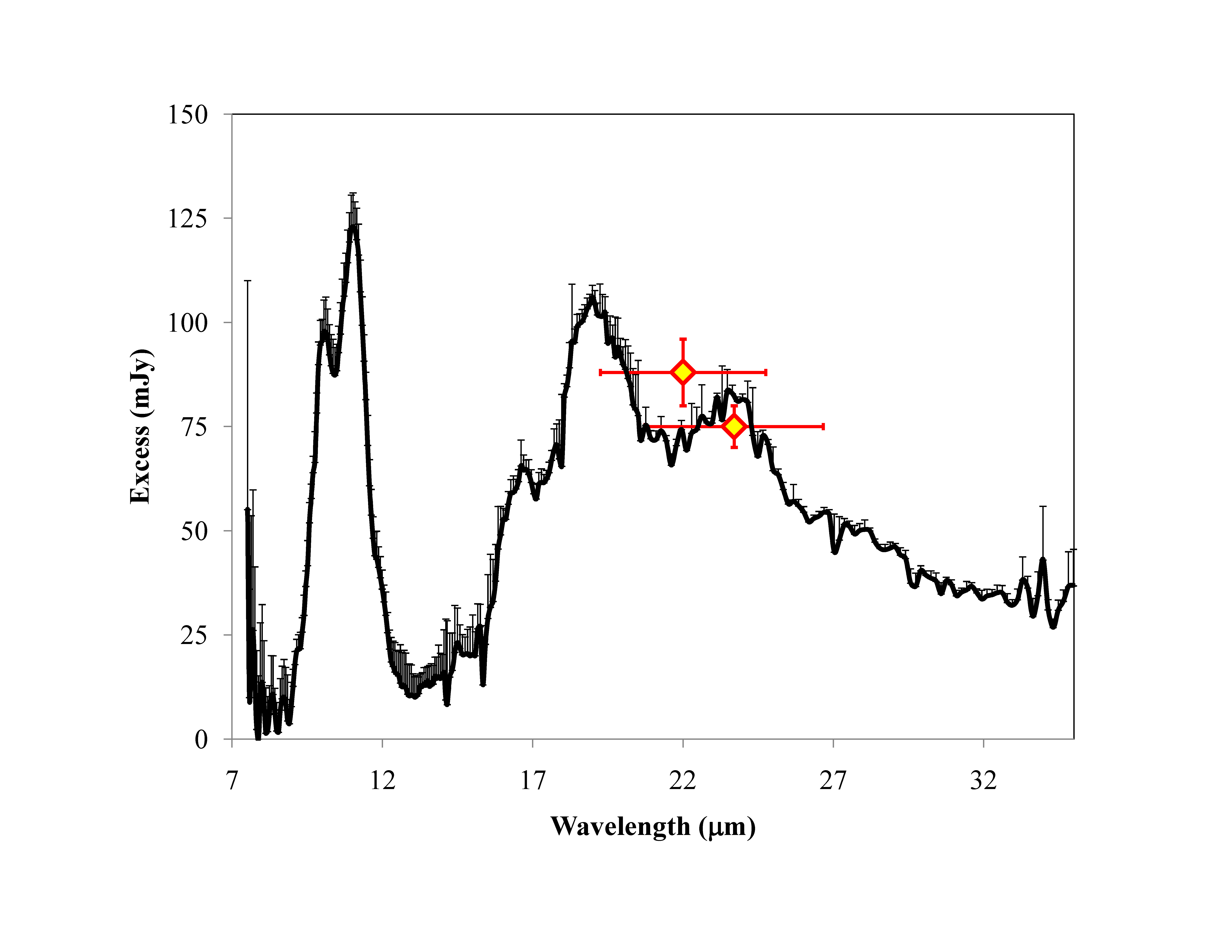} \\
\includegraphics[width=3.5in]{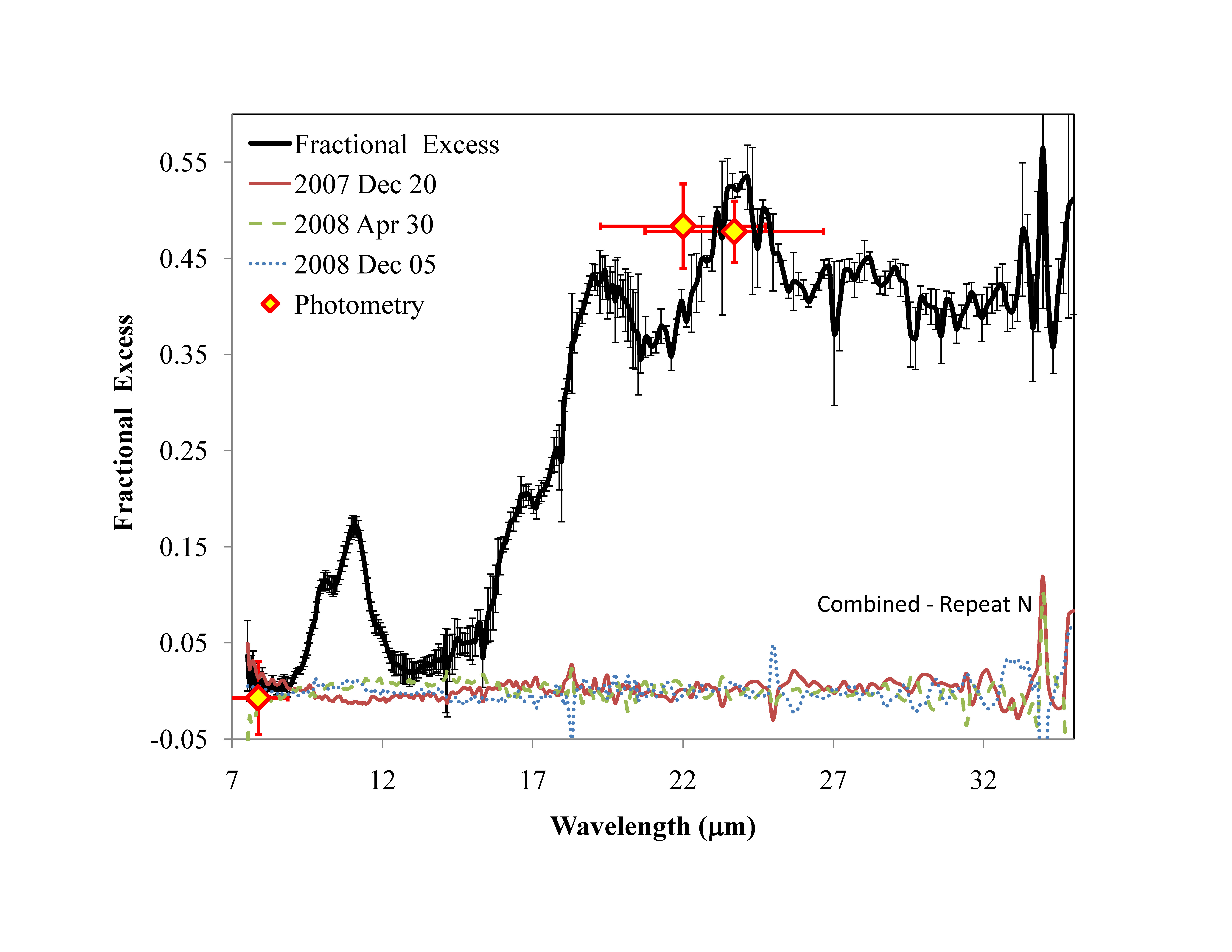}
\end{array}$
\end{center}
\caption{top) The low resolution spectrum of the excess of HD 69830 after subtraction of a Kurucz photospheric model. MIPS and IRS peakup photometry is shown for comparison. bottom) The solid black line shows the fractional excess in the LoRes spectrum of HD 69830 relative to the photosphere. The lower dotted lines show the differences between the {\it average} fractional excess observed over 1.4 year interval
vs. the individual 3 repeats making up the average, 2007 Dec (red solid), 2008 Apr (green dashed), and 
2008 Dec (blue dotted). There is little or no change at the 1\% level from epoch to the next. Variations due to
 instrumental noise are apparent at the short and long wavelength ends of the data. \label{HD 69830LoRes}}
\end{figure}


\clearpage
\begin{figure}[h]
\begin{center}$
\begin{array}{c}
\includegraphics[width=3.5in]{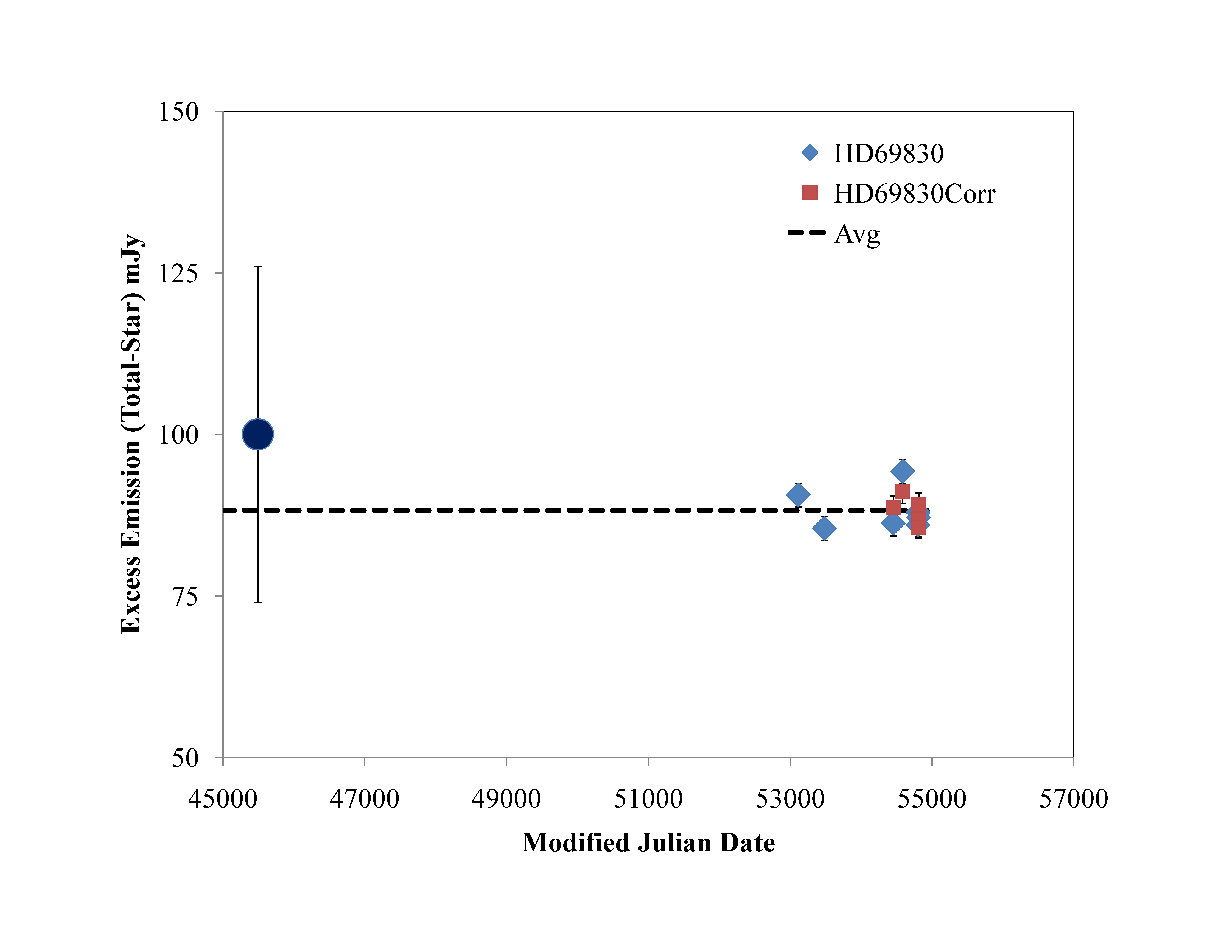} \\
\includegraphics[width=3.5in]{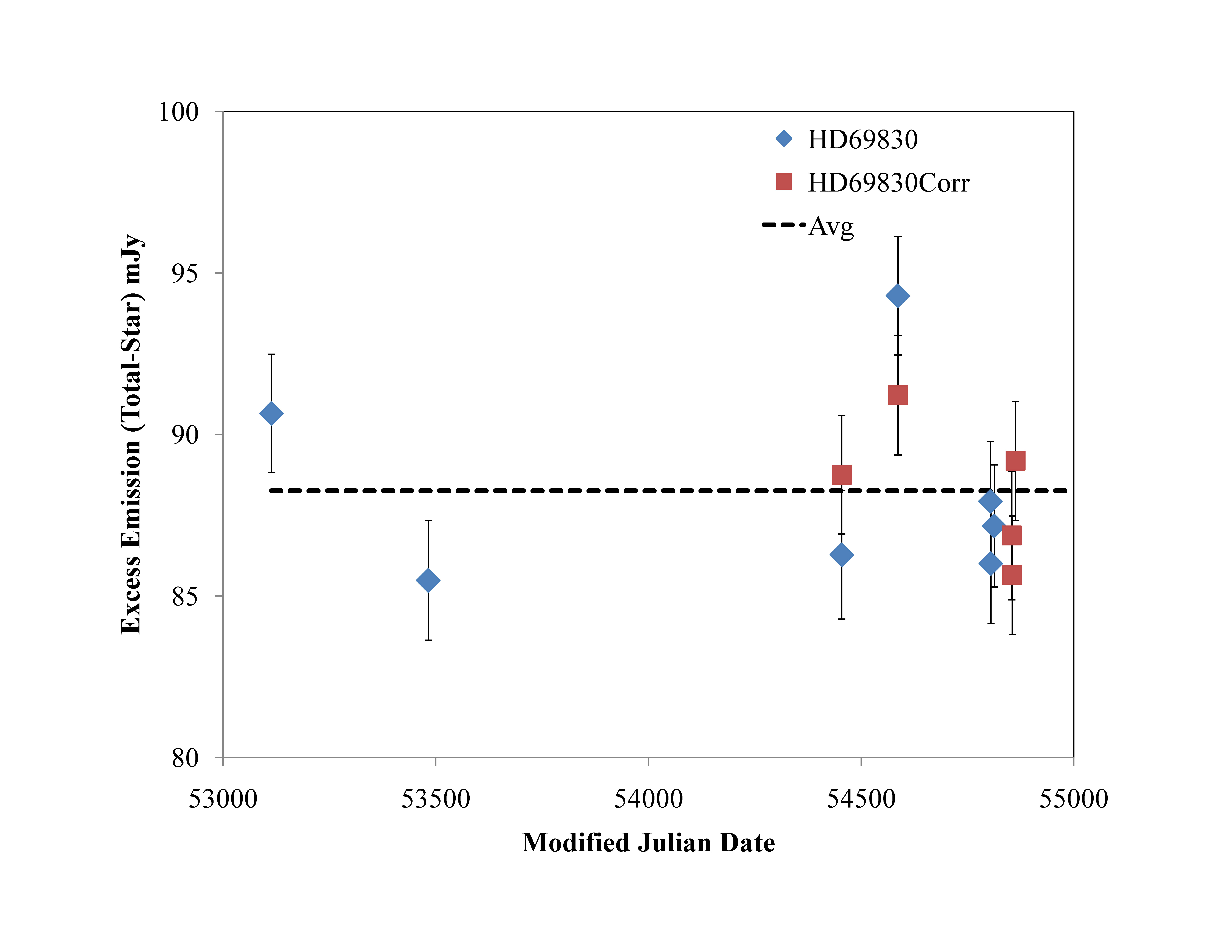}
\end{array}$
\end{center}
\caption{top) The excess emission from HD 69830 observed with the Spitzer/IRS peak-up array 
at 22 $\mu$m over 4 years (blue diamonds and red squares) is compared with an IRAS 25 $\mu$m datum from 24 years earlier (black circle).
Five IRS observations (red squares) have been corrected using contemporaneous peak-up imaging of HD68146;
Uncorrected flux values are shown as blue diamonds. The temporal variation of HD 69830's excess emission
is less than 33\% (1$\sigma$) over 24 years. bottom) Same as (a) but for Spitzer data only showing less than 3\% (1$\sigma$) variation over 4 years.\label{IRSPeakupFig}}
\end{figure}


\clearpage
\begin{figure}
\includegraphics[width=1.0\textwidth]{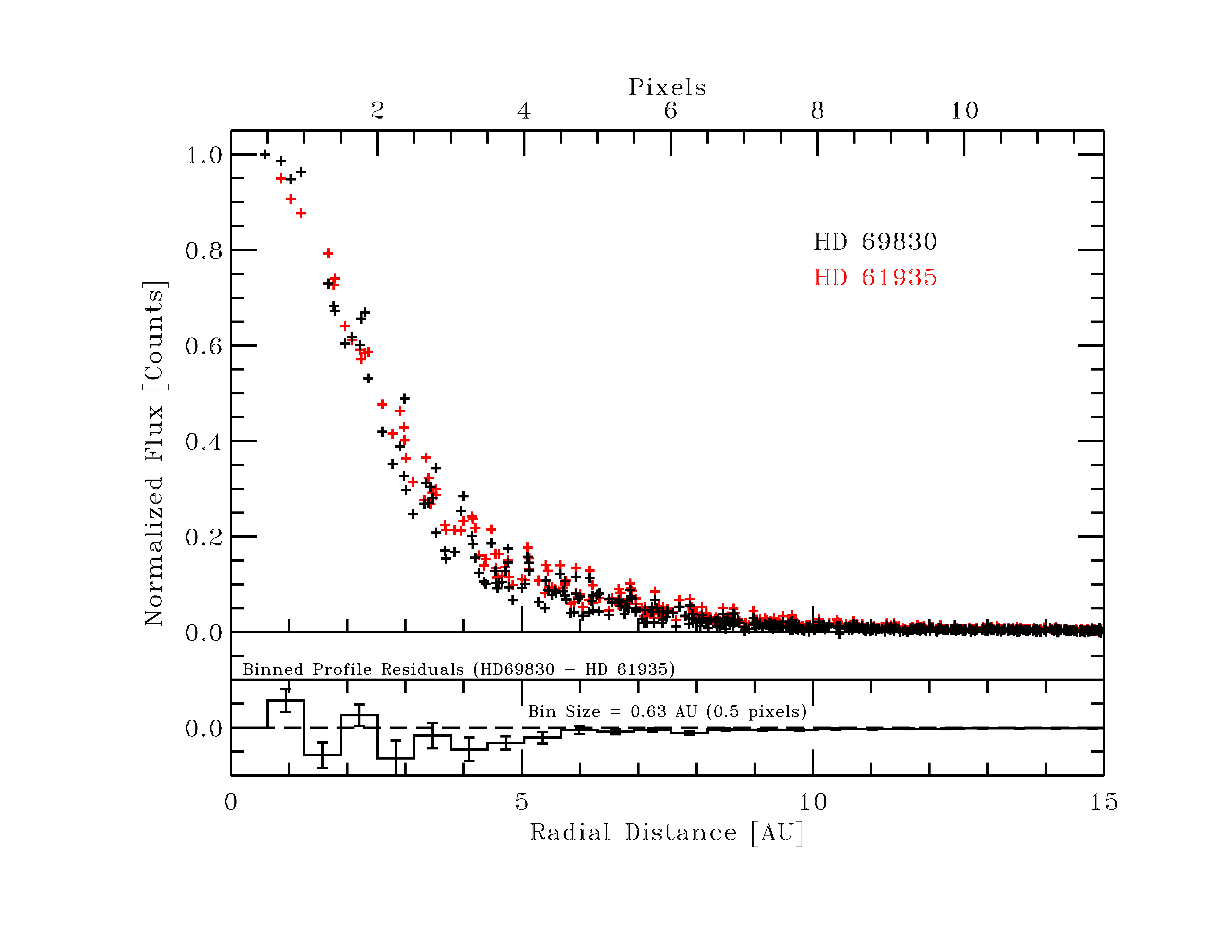}
\figcaption{{The azimuthally averaged 1-D profile of the 18.5 $\mu$m emission from
HD 69830 (black) measured with the Michelle instrument on the Gemini-N telescope is compared with a reference star, HD61935 (red). A upper limit to the size of the source is 0.3\arcsec\ (FWHM), or roughly 4 AU diameter.}
\label{Michelle}}
\end{figure}


\clearpage
\begin{figure}[tbp] 
 \centering
 \includegraphics[width=5.67in,height=7.33in,keepaspectratio]{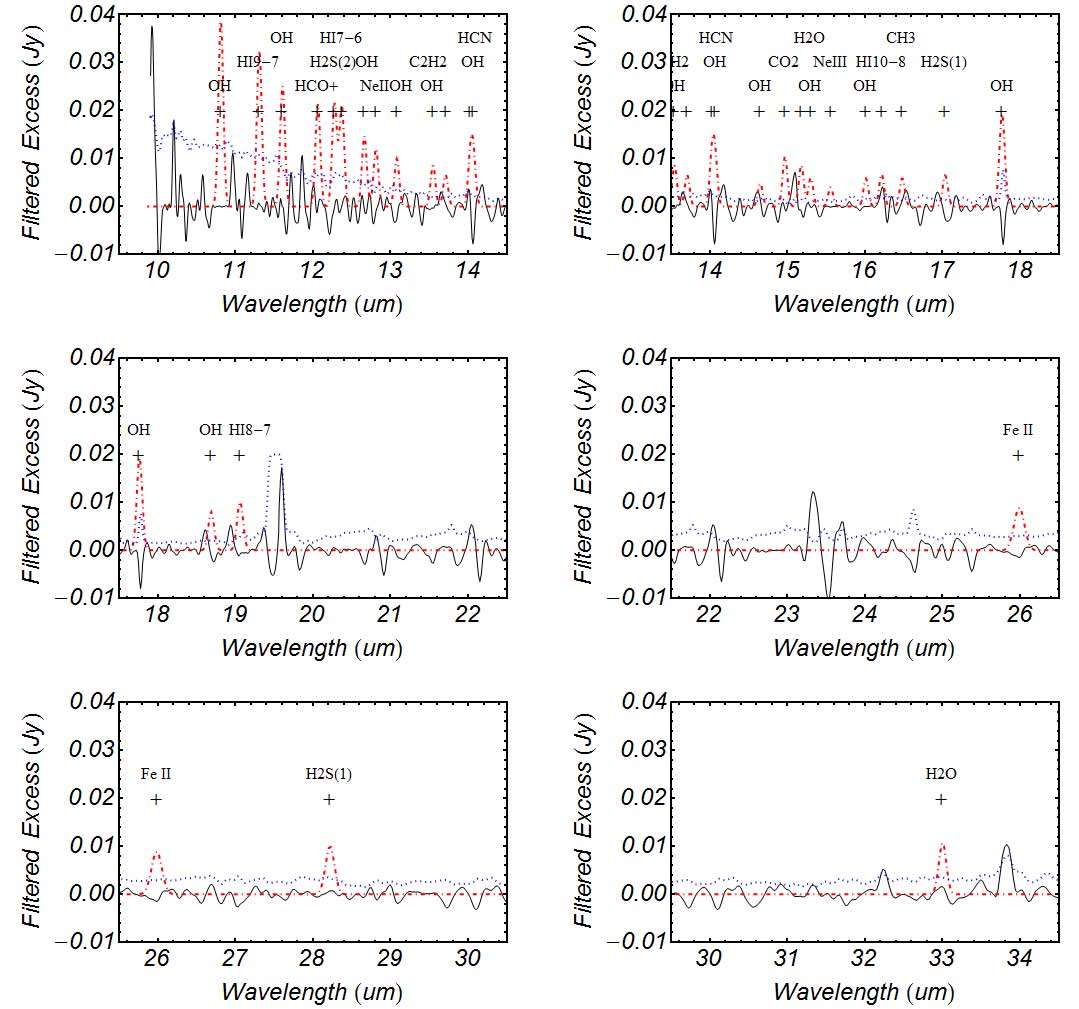}
 \caption{Each panel shows a portion of the smoothed, flattened spectrum of the excess from HD 69830. No spectral features
are identified with significance above the instrumental noise. The solid black line shows the data and the dashed red line shows what a spectral line at the level of the 3$\sigma$ upper limits would look like at the position of some of the lines discussed in the text. The blue dotted line represents the 1$\sigma$ noise level determined from the average of the multiple HiRes spectra. Other ``features" in the spectrum do not align with commonly expected gas species and can mostly can be traced back to instrumental artifacts (bad pixels, order boundaries, etc) in the data.}
 \label{FlattenHiresPanels}
\end{figure}


\clearpage
\begin{figure}
\centering
\includegraphics[width=\textwidth]{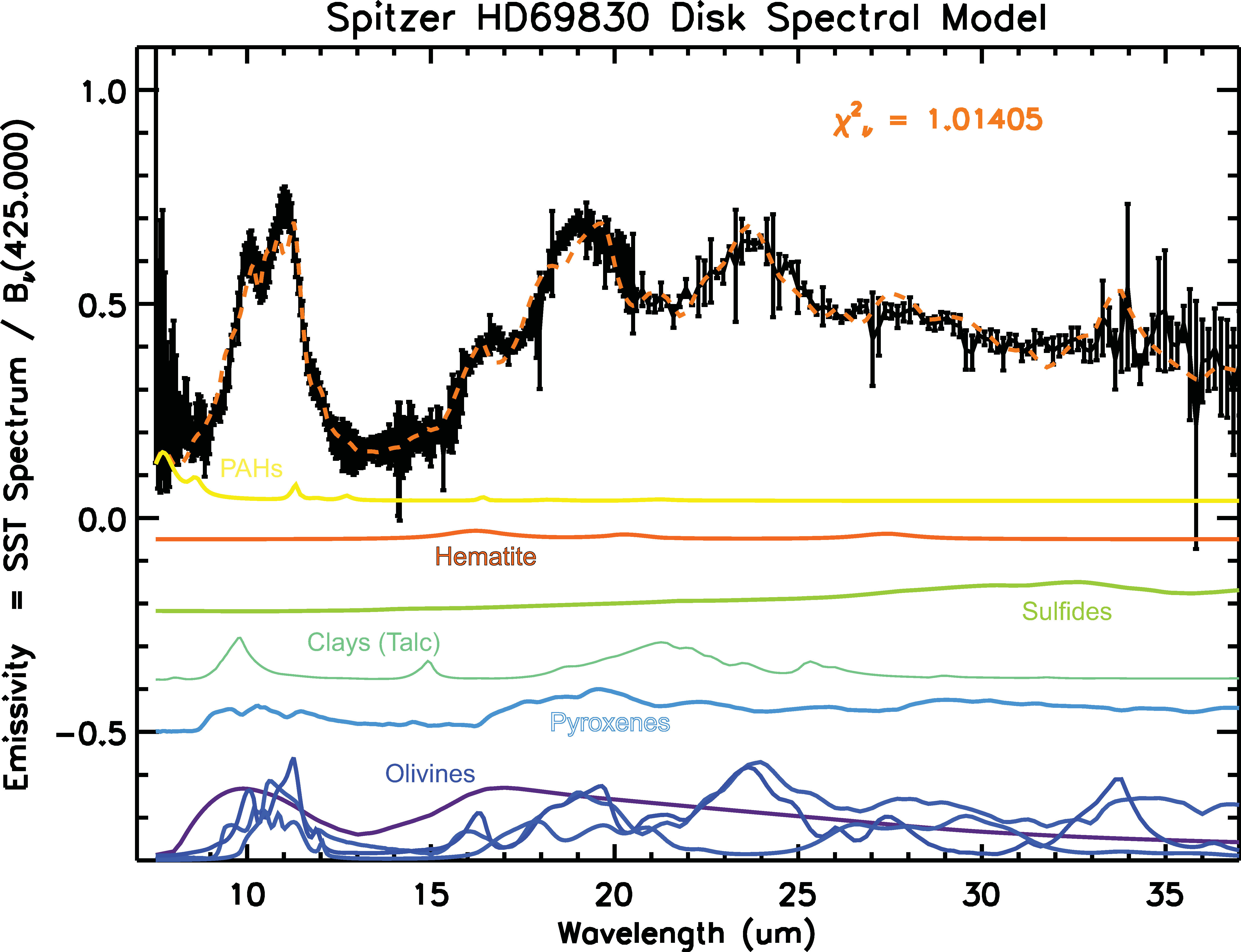}
 \caption{A decomposition of the HD 69830 spectrum reveals a composite of the mineralogical species listed in Table~\ref{caseytable}. The data (black) and the model fit (dotted orange line) are shown in terms of the excess emission divided by a 425 K blackbody to represent grain emissivities. Below the data and model curves are the emissivities of dominant species in the model. From top to bottom the species include: amorphous carbon (orange), sulfides (green), talc (yellow), pyroxenes (blue), and at the bottom a mixture of olivines (blue and purple).
}
\label{fit}
\end{figure}


\clearpage
\begin{figure}
\centering
\includegraphics[width=\textwidth]{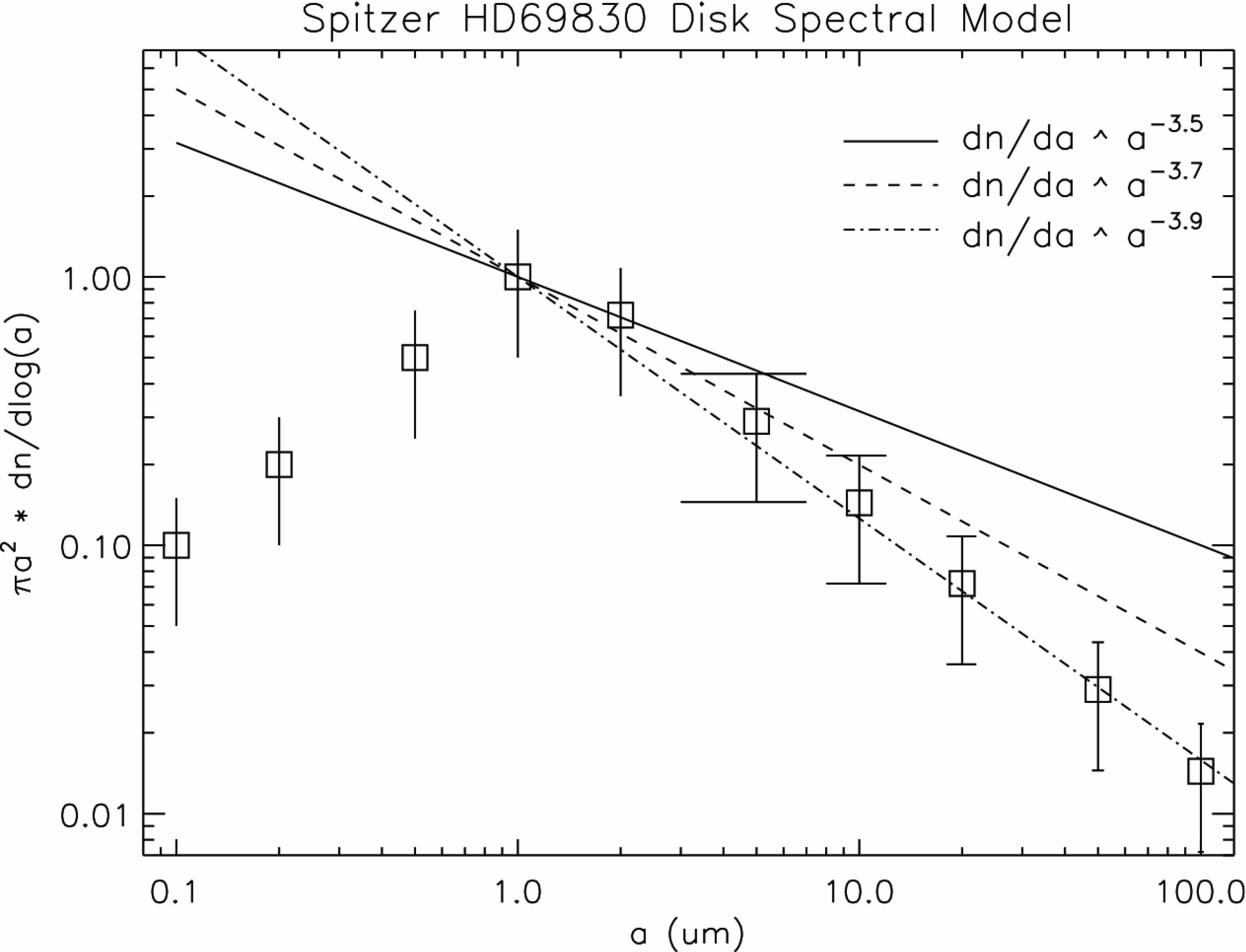}
 \caption{The particle size distribution for the grains responsible for the excess from HD 69830 follow a power-law distribution steeper than the canonical $-3.5$. 
}
\label{dustsize}
\end{figure}

\clearpage
\begin{figure}
\centering
\includegraphics[width=6in]{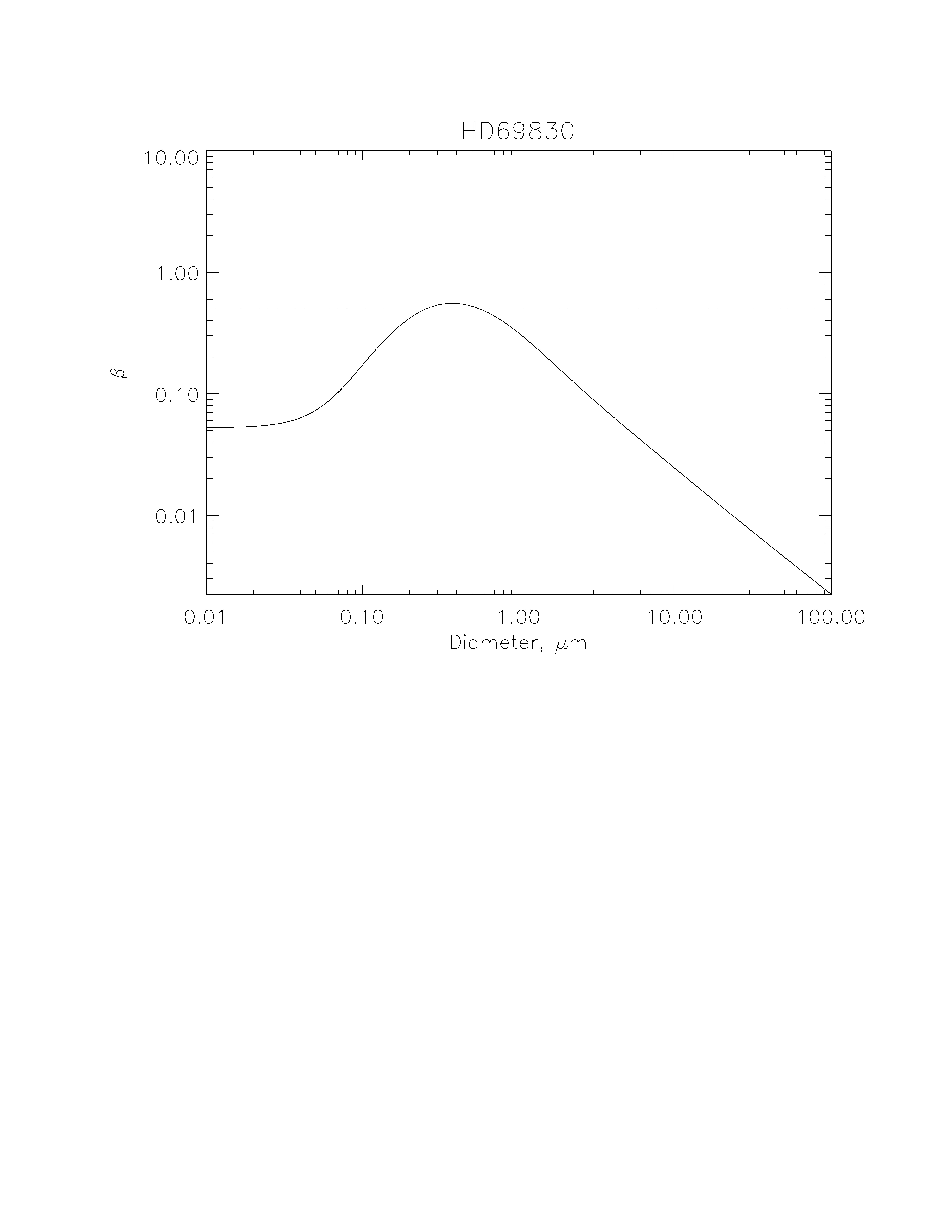}
 \caption{The factor $\beta= F_{rad}/F_{grav} $  represents the balance between radiative and gravitational forces for dust grains and depends on physical characteristics of the grains (size, composition) and the stellar radiation field \citep{wyatt1999, sheret04}. Calculations for silicate grains orbiting HD 69830 are shown. Only grains with  $\beta>0.5$, corresponding to a narrow range of radius,  will be blown out of the system.}
\label{betafig}
\end{figure}

\clearpage
\begin{figure}
\centering
\includegraphics[width=\textwidth]{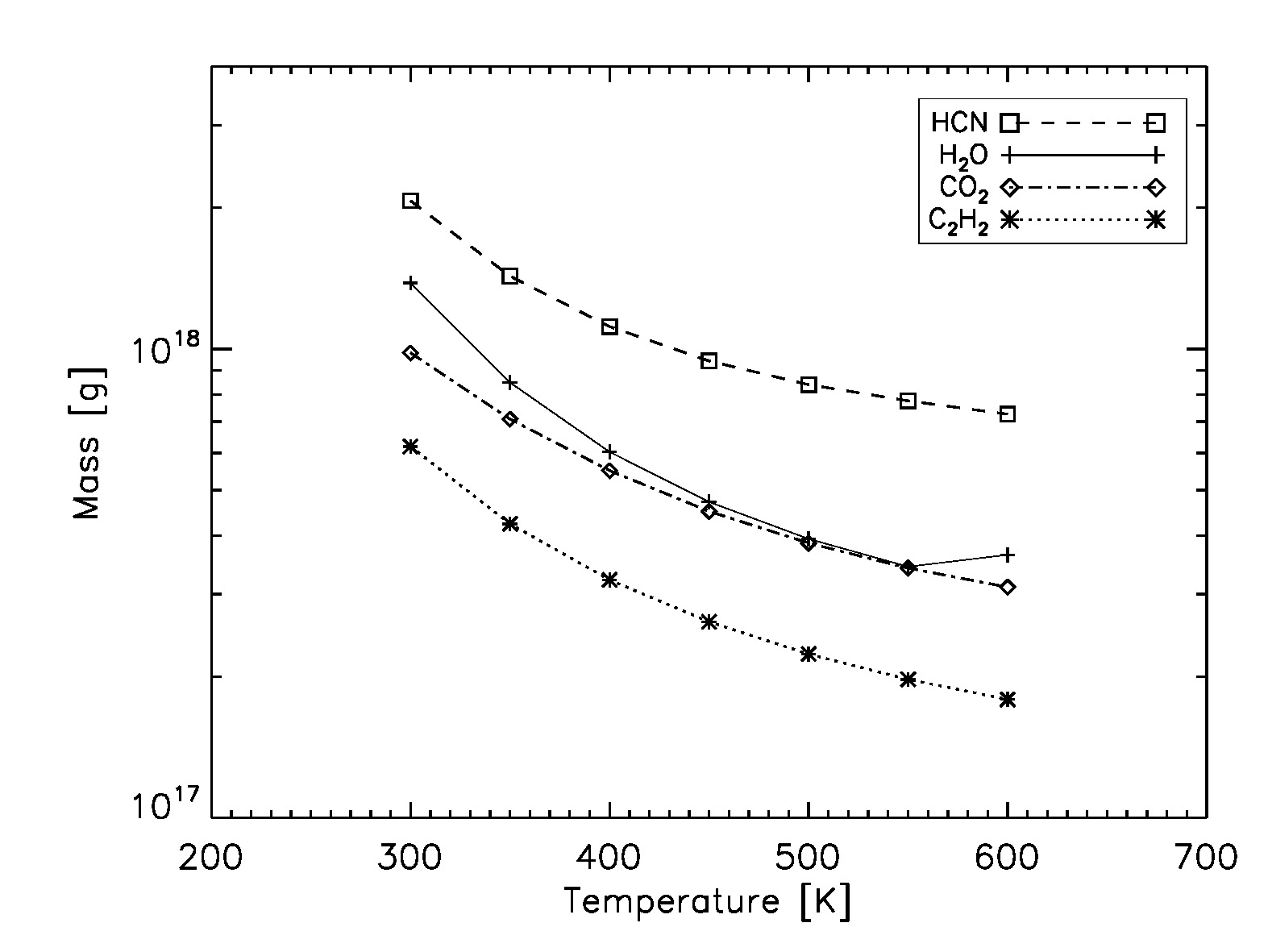}
 \caption{A simple optically thin model based on IRS HiRes spectroscopy sets limits to the amounts of different gaseous species that could be
present in the HD 69830 disk at various temperatures.
}
\label{gaslimit}
\end{figure}


\clearpage
\begin{figure}[tbp] 
 \centering
 \includegraphics[width=5.5in,height=7.12in,keepaspectratio]{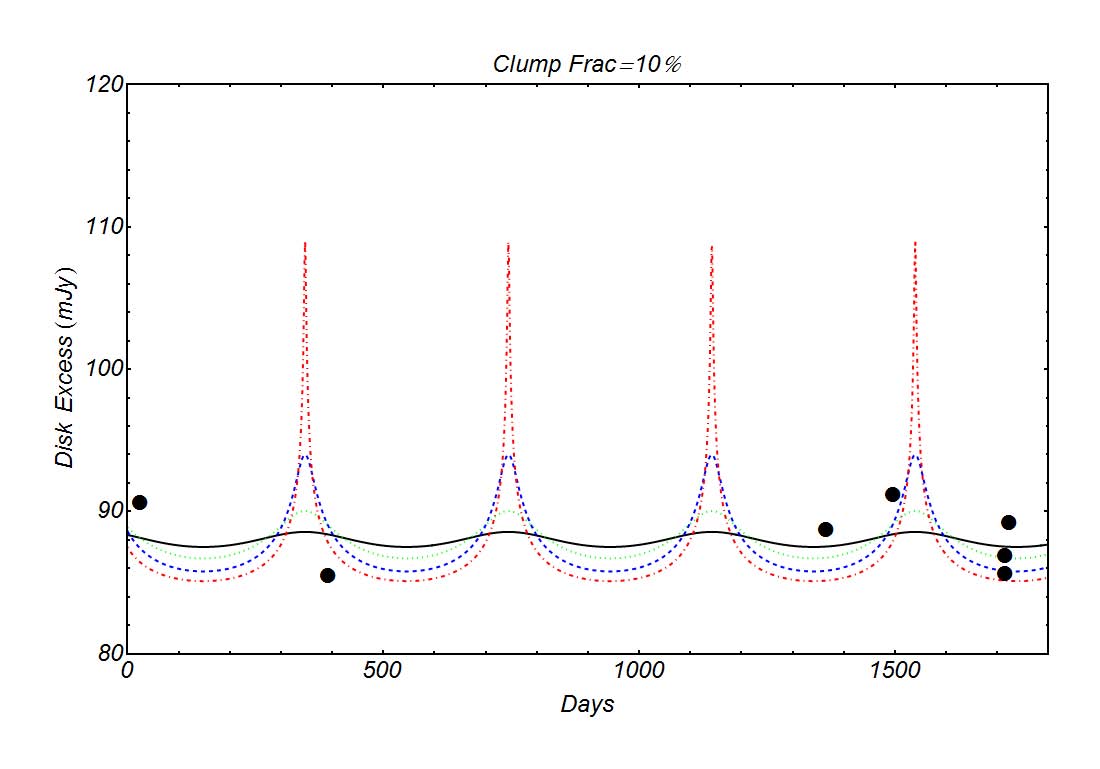}
\caption{A simple model shows how putting 10\% of the emitting material onto an eccentric orbit could modulate
the observed excess. The curves show the total excess including the variable emission from
dust with an orbital period of 400 days (1 AU semi major axis for HD 69830) and eccentricities of 0.1 (black solid line),0.3 (green dotted curve),
0.6 (blue dashed curve) and 0.9 (red, dash dotted curve) over approximately 4 years. 
The initial location of the clump is arbitrary. Spitzer/IRS points are shown for comparison.}
 \label{PlanckOrbit}
\end{figure}

\clearpage

\begin{figure}[tbp] 
 \centering
 \includegraphics[width=5.5in,height=7.12in,keepaspectratio]{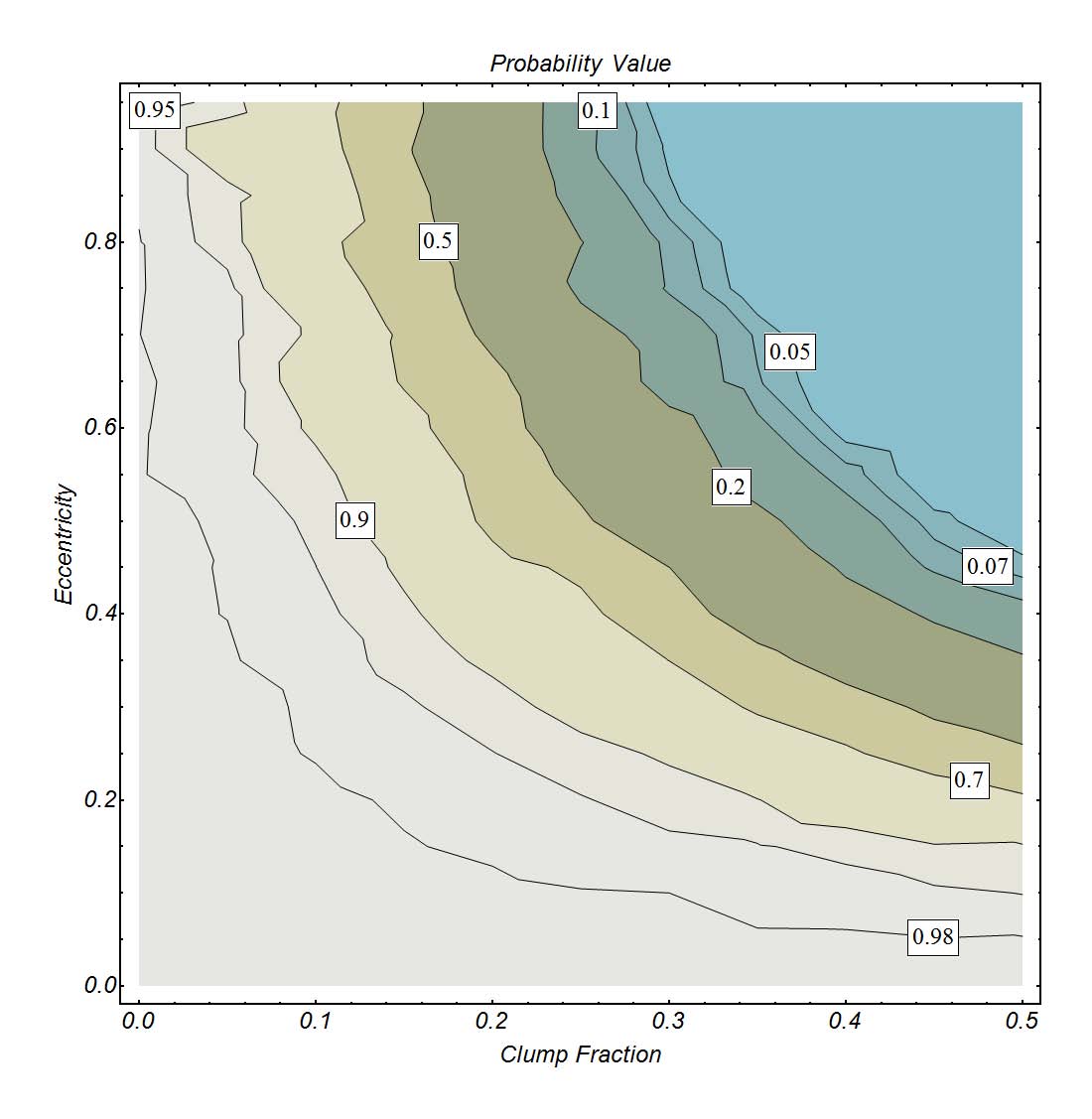}
\caption{The contours show $\chi^2$ probability values resulting from comparing the eccentric clump model with a simple model of constant emission using the IRS peak-up array measurements over 4 years. Only the most extreme values of clumpiness and eccentricity yield $p<0.05$ can be ruled out by our observations. Other combinations of clumpiness and eccentricity are consistent with the Spitzer/IRS data.}
 \label{PlanckOrbitFluxchi2}
\end{figure}

\clearpage
\begin{deluxetable}{rcccl} 
\tabletypesize{\scriptsize}
\tablecaption{ Observing Log \label{ObsLog}}
\tablehead{
\colhead{Obs ID}&\colhead{Star}&\colhead{Mode}&
 \colhead{Obs Date}&\colhead{Comment}}
\startdata
4016640&HD 69830&IRS/LoRes&2004 Apr 19&Repeat0, PID41\\
4041728&HD 69830&MIPS&2004 May 11&All wavelengths, PID41\\
12710656&HD 69830&IRS/HiRes&2005 Apr 22&No bkgnd/peak-up only, PID41\\
22341376&HD 69830&IRAC&2007 Nov 25&All wavelengths\\
22340608&HD 69830&MIPS& 2007 Nov 27& 24 \& 70 $\micron$ \\
22341888&HD 69830&IRS/LoRes&2007 Dec 20&Repeat2\\
22345728&HD 69830&IRS/HiRes&2007 Dec 20&Repeat2a\\
22346240&HD 69830&IRS/HiRes&2007 Dec 20&Repeat2b\\
22342144&HD 69830&IRS/LoRes&2008 Apr 30&Repeat3\\
22347776&HD 69830&IRS/HiRes&2008 Apr 30&Repeat3a\\
22348288&HD 69830&IRS/HiRes&2008 Apr 30&Repeat3b\\
22343680&HD 69830&IRS/HiRes&2008 Dec 04&Repeat1b\\
22340864&HD 69830&IRS/HiRes&2008 Dec 04&Repeat1a\\
22341632&HD 69830&IRS/LoRes&2008 Dec 05&Repeat1\\
28416000&HD 69830&IRS/HiRes&2008 Dec 13&Repeat4a\\
28830464&HD 69830&IRS/LoRes&2009 Jan 09&Repeat5, PID528 (DDT)\\
28830464&HD 69830&IRS/LoRes&2009 Jan 15&Repeat6, PID528 (DDT)\\ \hline
22344704&HD68146&IRS/HiRes&2007 Dec 20&Repeat2a\\
22345216&HD68146&IRS/HiRes&2007 Dec 20&Repeat2b\\
22342656&HD68146&IRS/LoRes&2007 Dec 20&Repeat2\\
22346752&HD68146&IRS/HiRes&2008 Apr 30&Repeat3a\\
22347264&HD68146&IRS/HiRes&2008 Apr 30&Repeat3b\\
22342912&HD68146&IRS/LoRes&2008 Apr 30&Repeat3\\
22343168&HD68146&IRS/HiRes&2008 Dec 04&Repeat1a\\
22344192&HD68146&IRS/HiRes&2008 Dec 04&Repeat1b\\
28415488&HD68146&IRS/HiRes&2008 Dec 13&Repeat4a\\
22342400&HD68146&IRS/LoRes&2008 Dec 05&Repeat1
\enddata
\
\tablecomments{All observations are from program PID40109 unless
otherwise stated. The numerology of ``repeats" represents names
associated with submitted AORs, not the order in which they were actually executed
by Spitzer. The background observation for HiRes Repeat3a was lost
due to downlink problems; Repeat 4a was added as a replacement. Gemini/Michelle
observations at 11.2 and 18.5 $\mu$m were made on 2007 Mar 7 and 8.
Keck Interferometer observations were taken at $K$ band on 2006 Nov 11
and at $L$ band on 2010 Feb 26.}

\end{deluxetable}

\clearpage
\begin{deluxetable}{lccccr}
\tablecaption{Infrared Photometry of HD 69830 \label{phottable}}
\tablehead{
Observing Mode & Wavelength & Observed $F_\nu$
&Photosphere $^a$ &Excess & $\chi$ $^b$ \\
& $(\mu$m)& (Jy)&(Jy) & (Jy) &
}
\startdata
Keck Interferometer & 2.2 & 14.0 $\pm$ 0.4$^c$ & 14.4 & -0.4 $\pm$ 0.4$^d$ & -1.0 \\
IRAC		&3.55&6.345 $\pm$ 0.18	&6.300&0.045 $\pm$ 0.191&0.2\\ 
Keck Interferometer & 3.91 & 5.2 $\pm$ 0.2$^c$ & 5.2 & 0.0 $\pm$ 0.2$^e$ & 0.0 \\
IRAC		&4.49&4.028 $\pm$ 0.12	&3.756&0.272 $\pm$ 0.126 &2.2 \\ 
IRAC		&5.73&2.604 $\pm$ 0.06	&2.491&0.113 $\pm$ 0.065 & 1.7 \\ 
IRAC		&7.87&1.370 $\pm$ 0.05	&1.381&-0.011 $\pm$ 0.052 &-0.2\\ 
Gemini-Michelle	&11.2&0.76 $\pm$ 0.07$^f$ &0.572&0.18 $\pm$ 0.07 &2.5 \\
IRAS		&12.0&0.68 $\pm$ 0.04	&0.610&0.070 $\pm$ 0.041 & 1.7 \\ 
Gemini-Michelle	&18.5&0.61 $\pm$ 0.19$^f$ &0.258&0.35 $\pm$ 0.20 &1.7 \\
IRS Peak-up	&22.0&0.270 $\pm$ 0.008	&0.182&0.088 $\pm$ 0.008 & 11 \\
MIPS (2004)	&23.7&0.232 $\pm$ 0.005	&0.157&0.075 $\pm$ 0.005 & 15 \\
MIPS (2007)	&23.7&0.235 $\pm$ 0.005	&0.157&0.078 $\pm$ 0.005 & 15 \\
IRAS		&25.0&0.24 $\pm$ 0.026	&0.141&0.100 $\pm$ 0.026 &3.8\\
MIPS (2004)$^g$	&71.4&0.019 $\pm$ 0.004	&0.018&0.001 $\pm$ 0.004 &2.6\\ 
MIPS (2007)	&71.4&0.015 $\pm$ 0.003	&0.018&-0.003 $\pm$ 0.003 &-1.0\\ 
\enddata
\tablenotetext{a}{From Kurucz model fitted to data from visible to 4.5 $\mu$m with adopted fractional error in the model of 1\%. Overall fit of model to data has reduced $\chi^2$ of 1.29.} 
\tablenotetext{b}{ $\chi$ =(Observed- Photosphere)/$\sqrt{\rm noise^2+(model \, uncertainty)^2}$ where IRAC calibration uncertainties are taken to be 3\%,
peak-up uncertainty is set to 3\%, and MIPS 24 $\mu$m uncertainty is 2\% \citep{engelbracht07}. MIPS 70 $\mu$m uncertainty is dominated by the background noise 
in the image, rather than by systematics. We adopt the MIPS 2007 observation as being of much higher quality in terms of understanding of instrument performance, improved pointing and much greater integration time, 300 vs 2,000 sec. The noise values quoted in the table are total noise values, including statistical and confusion noise as well as calibration uncertainties.}
\tablenotetext{c}
{Keck Interferometer Nuller measurements are relative to the stellar photosphere. 
The fluxes here are scaled from the given photospheric flux.}
\tablenotetext{d}{The KI K-band limit applies to dust within 0.6 AU.}
\tablenotetext{e}{The KI L-band limit applies to dust within 1.1 AU.}
\tablenotetext{f}{In a 2\arcsec\ diameter aperture.}
\tablenotetext{g}{As reported in Bryden et al (2006). When re-reduced with same software pipeline as the 2007 data, the revised 2004 value is 0.026$\pm$0.003 Jy which is still consistent with no excess. The 2004 data are of considerably lower quality than the 2007 data.}
\end{deluxetable}

\clearpage
\begin{deluxetable}{lcccccc} 
\tablecaption{IRS Peak-Up Data \label{IRSPeakup}}
\tablehead{
\colhead{Star}	& Date & Peak-Ups & Frames & Raw $F_{\nu}$$^a$ &Corrected $F_{\nu}$$^b$&Excess$^c$\\
	& & & &(mJy)&(mJy)&(mJy)
}
\startdata
HD 69830	&2004 Apr 19 & 1 & 6 &272.8$\pm$0.2& &90.7$\pm$1.8\\
HD 69830	&2005 Apr 22 & 1 & 6 &267.7$\pm$0.3& &85.5$\pm$1.8\\
HD 69830	&2007 Dec 20 & 3 & 18 &268.4$\pm$0.8&270.9$\pm$0.8&88.8$\pm$2.0\\
HD 69830	&2008 Apr 30 & 3 & 17$^d$ &276.5$\pm$0.2&273.4$\pm$0.2&91.2$\pm$1.8\\
HD 69830	&2008 Dec 04 & 2 & 12 &270.1$\pm$0.3&269.0$\pm$0.3&86.9$\pm$2.8 \\
HD 69830	&2008 Dec 05 & 1 & 6 &268.2$\pm$0.4&267.8$\pm$0.4&85.6$\pm$1.9\\
HD 69830	&2008 Dec 13 & 1 & 6 &269.3$\pm$0.5&271.4$\pm$0.5&89.2$\pm$1.9\\
HD 69830	&2009 Jan 09 & 1 & 6 &269.6$\pm$0.4 & &87.4$\pm$1.9\\
HD 69830	&2009 Jan 15 & 1 & 6 &261.6$\pm$0.4 & &79.4$\pm$1.9\\
HD 69830 & Average	& 14 & 83 &270.3$\pm$1.2&270.3$\pm$1.1&88.1$\pm$1.8 \\
\hline
&&&&&Normalization\\
HD68146&2007 Dec 20 & 3 & 17$^d$ &153.6$\pm$0.6&0.991&\\
HD68146&2008 Apr 30 & 3 & 18 &156.8$\pm$0.2&1.011&\\
HD68146&2008 Dec 04 & 2 & 12 &155.6$\pm$0.3&1.004&\\
HD68146&2008 Dec 05 & 1 & 6 &155.2$\pm$0.3&1.001&\\
HD68146&2008 Dec 13 & 1 & 6 &153.9$\pm$0.6&0.993&\\
HD68146&Average& 10 & 59 &155.0$\pm$0.6& &
\enddata
\tablenotetext{a}{Stated uncertainties come from the standard deviation of the mean for all points in the dataset 
(i.e. the deviation between the frames at each epoch or between the epochs for the overall average).}
\tablenotetext{b}{Corrected using contemporaneous observations of HD68146.}
\tablenotetext{c}{Excess relative to photospheric flux density of 182.2 mJy at 22.0 $\mu$m. 
Uncertainty for the excess includes 1\% for the subtracted stellar model. 
Overall calibration uncertainty of $\sim$3\% is not included here.}
\tablenotetext{d}{One corrupted frame removed.}
\end{deluxetable}

\clearpage
\begin{deluxetable}{lcccc} 
\tablecaption{IRS Low Resolution Spectrum (Electronic Table) \label{IRSLRSdata}}
\tablehead{Wavelength&Flux$^a$&Photosphere&Excess&FracExcess\\
	($\mu$m)&(mJy)&(mJy)&(mJy)&\\}
\startdata
7.516 &1560.0$\pm$55.0 &1504.9& 55.1$\pm$55.0 &0.0366$\pm$0.0365\\
7.576&1493.2$\pm$26.0&1483.3&10.0$\pm$26.0&0.0067$\pm$0.0175\\
7.637&1480.0$\pm$33.2&1459.6&20.4$\pm$33.2&0.0140$\pm$0.0228\\
7.697&1463.2$\pm$33.7&1437.1&26.1$\pm$33.7&0.0182$\pm$0.0235\\
7.758&1429.9$\pm$27.6&1416.2&13.7$\pm$27.6&0.0097$\pm$0.0195\\
7.818&1397.4$\pm$18.9&1395.1&2.4$\pm$18.9&0.0017$\pm$0.0136\\
7.879&1374.7$\pm$15.3&1374.9&-0.1$\pm$15.3&-0.0001$\pm$0.0132\\
7.939&1364.8$\pm$17.6&1354.4&10.4$\pm$17.6&0.0076$\pm$0.0129\\
8.000&1347.7$\pm$18.7&1334.1&13.6$\pm$18.7&0.0102$\pm$0.0140
\enddata
\tablenotetext{a}{Stated uncertainties are the larger of 1) standard deviation of the mean of three individual repeats or 2)
the average of the three noise values from the individual repeats divided by $\sqrt{3-1}$. No systematic error for
the subtraction of the photospheric model has been applied.}
\end{deluxetable}

\clearpage
\begin{deluxetable}{lcccc} 
\tabletypesize{\scriptsize}
\tablecaption{Flux Limits for Selected Lines in the HiRes Spectrum \label{HiResLines}}
\tablehead{
Line	& Wavelength & Amplitude $^a$ & FWHM $^b$& Line Flux Limits$^a$ \\
	& ($\mu$m)& (mJy)& ($\mu$m) & (3 $\sigma$, 10$^{-17}$ W m$^{-2}$)}
\startdata
PAH   & 11.2 &   33.1 &  0.076 & 10.59\cr 
PAH   & 12.0 &   23.2 &  0.075 &  6.45\cr 
PAH   & 12.7 &   15.0 &  0.075 &  3.71\cr 
PAH   & 16.4 &   7.9 &  0.078 &  1.23\cr 
PAH   & 17.4 &   4.0 &  0.082 &  0.58\cr 
PAH   & 17.7 &   8.1 &  0.084 &  1.16\cr\hline 
OH   & 10.815 &   38.8 &  0.076 & 13.34\cr 
HI 9-7  & 11.310 &   32.6 &  0.075 & 10.21\cr 
OH   & 11.609 &   25.1 &  0.075 &  7.46\cr 
HCO+  & 12.064 &   21.4 &  0.075 &  5.87\cr 
H$_2$ S(2)  & 12.279 &   21.4 &  0.075 &  5.66\cr 
HI 7-6  & 12.373 &   20.9 &  0.075 &  5.44\cr 
OH   & 12.663 &   14.7 &  0.075 &  3.65\cr 
NeII  & 12.813 &   12.0 &  0.075 &  2.92\cr 
OH   & 13.081 &   9.9 &  0.075 &  2.31\cr 
OH   & 13.547 &   8.5 &  0.075 &  1.85\cr 
C$_2$H$_2$  & 13.710 &   6.6 &  0.075 &  1.40\cr 
HCN   & 14.030 &   9.8 &  0.075 &  1.99\cr 
OH   & 14.070 &   8.8 &  0.075 &  1.77\cr 
OH   & 14.651 &   4.8 &  0.075 &  0.90\cr 
CO$_2$   & 14.970 &   10.5 &  0.076 &  1.88\cr 
H$_2$O   & 15.180 &   8.7 &  0.076 &  1.52\cr 
OH   & 15.302 &   6.0 &  0.076 &  1.03\cr 
NeIII  & 15.558 &   4.0 &  0.076 &  0.68\cr 
OH   & 16.013 &   5.8 &  0.077 &  0.93\cr 
HI 10-8  & 16.215 &   6.7 &  0.078 &  1.06\cr 
CH$_3$   & 16.482 &   5.9 &  0.079 &  0.91\cr 
H$_2$ S(1)  & 17.035 &   6.5 &  0.081 &  0.96\cr 
OH   & 17.767 &   19.3 &  0.084 &  2.75\cr 
OH   & 18.690 &   8.0 &  0.091 &  1.11\cr 
HI 8-7  & 19.062 &   10.2 &  0.094 &  1.40\cr 
Fe II  & 25.988 &   8.9 &  0.159 &  1.12\cr 
H$_2$ S(1)  & 28.221 &   10.0 &  0.138 &  0.92\cr 
H$_2$O   & 33.000 &   10.9 &  0.114 &  0.60\cr 
SiII  & 34.814 &   13.2 &  0.128 &  0.74\cr 
\enddata
\tablenotetext{a}{3$\sigma$ upper limit.}
\tablenotetext{b}{Typical spectral resolution based on the examination of spectral lines
observed in the shell star HD142926 and reduced in identical fashion.}
\end{deluxetable}

\clearpage
\begin{deluxetable}{lccccccc}
\tabletypesize{\scriptsize}
\tablecaption{
Composition of the Best-Fit Model of the HD 69830 Spectra
\label{caseytable}}

\tablehead{
Species & Weighted$^a$ & Density & Molecular & $N_{moles}$$^b$ & 
$T_{max}$ & Model $\chi^2$ $^c$ \\
& Surface Area & (g cm$^{-3}$) & Weight & (relative) & (K) & if not included}
\startdata
\multicolumn{7}{c}{\it Detections} \\
\multicolumn{1}{c}{\it Olivines} \\
Amorph Olivine (MgFeSiO$_4$) & 0.15 & 3.6 & 172 & 0.31 & 385 &10.4 \\
ForsteriteKoike (Mg$_2$SiO$_4$) & 0.26 & 3.2 & 140 & 0.59 & 385 &8.64 \\
Forsterite038 (Mg$_2$SiO$_4$) $^d$ & 0.21 & 3.2 & 140 & 0.48 & 385 &8.82 \\
Fayalite (Fe$_2$SiO$_4$) &0.12 &4.3 & 204 &0.25& 385& 3.85 \\\hline

\multicolumn{1}{c}{\it Pyroxenes} \\
Bronzite (Mg$_{1-x}$Fe$_x$Si$_2$O$_6$) $^d$ & 0.11 & 3.5 & 232 & 0.17 & 385 & 4.25 \\\hline

\multicolumn{1}{c}{\it Fe Species} \\
Sulfides (Fe$_{90}$Mg$_{10}$S)& 0.08 &4.5& 72 &0.50& 385 & 2.33 \\
Oxides (Fe$_3$O$_4$) $^{d,e}$ &0.02 &5.2 &232 &0.04& 385& 1.19 \\ \hline
\\
\\
\multicolumn{7}{c}{\it Marginal or Non-Detections} \\
\multicolumn{1}{c}{\it Phyllosilcates} \\
Talc (Mg$_3$Si$_4$O$_{10}$(OH)$_2$) $^e$& 0.05& 2.8& 379 &0.04 &385 &1.22 \\ \hline
\multicolumn{1}{c}{\it Organics} \\
Amorphous Carbon (C) &$<$0.005 &2.5 &12 &$<$0.1 &425 &1.03 \\ 
PAH (C$_{10}$H$_{14}$), ionized & $<$0.12 &1.0 & $<$178$>$ & $<$0.07 &N/A &1.10 \\\hline
\multicolumn{1}{c}{\it Water} \\
Water Ice $^f$ & $<$0.05 & 1.0 &18 &$<$0.28 &170 & 1.05\\ \hline
\multicolumn{1}{c}{\it Carbonates} \\
Magnesite (MgCO$_3$) &$<$0.02 &3.1 &84 &$<$0.07 &385 &1.03 \\
Dolomite (CaMgC$_2$O$_6$) $^{d,f}$ & $<$0.005 &2.9 &184 &$<$0.008 &385 &1.01 \\
Siderite(CaMgC$_2$O$_6$) d,f  &$<$0.01& 3.9& 116& $<$0.03&      385&    1.01\\ \hline
\multicolumn{1}{c}{\it Pyroxenes} \\
FerroSilite (Fe$_2$Si$_2$O$_6$) $^f$ &$<$0.005& 4.0 &264 &$<$0.008& 355 &1.01 \\
Diopside (CaMgSi$_2$O$_6$) $^f$ & $<$0.005 &3.3 &216 &$<$0.008& 385 &1.01 \\
\enddata
\tablenotetext{a}{Weight of the emissivity spectrum of each dust species required to match HD 69830.}
\tablenotetext{b}{$N_{moles} \propto$ Weighted Surface Area*Density/Molecular Weight. Errors are 10\% (2 $\sigma$).}
\tablenotetext{c}{Total best fit model $\chi^2$ = 1.03. Assumes 4\% absolute calibration uncertainty.}
\tablenotetext{d}{Not found in cometary systems to date.}
\tablenotetext{e}{New minor species detection not found in \citet{lisse07}.}
\tablenotetext{f}{Minor species reported detected in \citet{lisse07} but not found in higher SNR spectrum.}
\end{deluxetable}

\clearpage
\begin{deluxetable}{lcc}
\tablecaption{Maximum Molecular Masses From IRS Spectroscopy (400 K) \label{colette}}
\tablehead{
Molecule & Maximum Mass (g) & $\lambda (\mu$) \\}
\startdata
C$_2$H$_2$ & 6.5E+16 &  13.71\\
  CO$_2$ & 1.5E+17 &  14.97\\
  H$_2$O & 1.9E+17 &  33.00\\
  HCN & 2.4E+17 &  14.03\\
   OH & 3.6E+19 &  18.69\\
\enddata
\tablecomments{Models assume LTE excitation at T=400 K and a Gaussian 
local line broadening with $\sigma=2$ km/s. $\lambda$ gives the wavelength 
of the strongest spectral feature.}
\end{deluxetable}

\end{document}